\documentclass[lettersize,journal]{IEEEtran}
\usepackage{amsmath,amsfonts}
\usepackage{algorithmic}
\usepackage{algorithm}
\usepackage{array}
\usepackage[caption=false,font=normalsize,labelfont=sf,textfont=sf]{subfig}
\usepackage{textcomp}
\usepackage{stfloats}
\usepackage{url}
\usepackage{verbatim}
\usepackage{graphicx}
\usepackage{cite}
\usepackage{multirow}
\usepackage{adjustbox}
\usepackage{makecell}
\usepackage{mathtools}
\usepackage{comment}
\usepackage[normalem]{ulem}
\usepackage[inline]{enumitem}
\usepackage{soul}
\usepackage{bm}
\usepackage{siunitx}
\usepackage{algorithm}
\usepackage{subfig}
\usepackage{amssymb}
\usepackage[font=small]{caption}
\usepackage{xr}
\makeatletter
\newcommand*{\addFileDependency}[1]{
  \typeout{(#1)}
  \@addtofilelist{#1}
  \IfFileExists{#1}{}{\typeout{No file #1.}}
}
\makeatother

\setstcolor{red}

\newtheorem{lemma}{\textbf{Lemma}}
\newtheorem{theorem}{\textbf{Theorem}}

\newcolumntype{L}[1]{>{\raggedright\arraybackslash}m{#1}}
\newcolumntype{C}[1]{>{\centering\arraybackslash}m{#1}}

\begin{document}

\title{Quantum-Assisted Online Task Offloading and Resource Allocation in MEC-Enabled Satellite-Aerial-Terrestrial Integrated Networks}

\author{Yu~Zhang,~\IEEEmembership{Student~Member,~IEEE,}
        Yanmin~Gong,~\IEEEmembership{Senior~Member,~IEEE,}
        Lei~Fan,~\IEEEmembership{Senior~Member,~IEEE,}
        Yu~Wang,~\IEEEmembership{Fellow,~IEEE,}
        Zhu~Han,~\IEEEmembership{Fellow,~IEEE}, and Yuanxiong~Guo,~\IEEEmembership{Senior~Member,~IEEE}
\thanks{Y. Zhang and Y. Gong are with the Department of Electrical and Computer Engineering, University of Texas at San Antonio, Texas, 78249, USA. (e-mail: \{yu.zhang@my., yanmin.gong@\}utsa.edu).}
\thanks{L. Fan is with the Department of Engineering Technology and Department of Electrical and Computer Engineering, University of Houston, Houston, TX 77204 USA (e-mail: lfan8@central.uh.edu).}
\thanks{Y. Wang is with Department of Computer and Information Sciences, Temple University, Philadelphia, Pennsylvania 19122, USA. (e-mail: wangyu@temple.edu).}
\thanks{Z. Han is with the Department of Electrical and Computer Engineering, University of Houston, Houston, TX 77004 USA, and also with the Department of Computer Science and Engineering, Kyung Hee University, Seoul 446-701, South Korea. (e-mail: zhan2@uh.edu).}
\thanks{Y. Guo is with the Department of Information Systems and Cyber Security, University of Texas at San Antonio, Texas, 78249, USA. (e-mail: yuanxiong.guo@utsa.edu).}
\thanks{The work is partially supported by the US NSF (Grant NO. CNS-2047761, CNS-2106761, CNS-2318683, CMMI-2222670, CNS-2006604, CNS-2128378, OAC-2417716, ECCS-2045978, ECCS-2302469.), Toyota, Amazon, and Japan Science and Technology Agency (JST) Adopting Sustainable Partnerships for Innovative Research Ecosystem (ASPIRE) under JPMJAP2326. (Corresponding author: Y. Guo.)}}



\maketitle
\begin{abstract}
In the era of Internet of Things (IoT), multi-access edge computing (MEC)-enabled satellite-aerial-terrestrial integrated network (SATIN) has emerged as a promising technology to provide massive IoT devices with seamless and reliable communication and computation services. This paper investigates the cooperation of low Earth orbit (LEO) satellites, high altitude platforms (HAPs), and terrestrial base stations (BSs) to provide relaying and computation services for vastly distributed IoT devices. Considering the uncertainty in dynamic SATIN systems, we formulate a stochastic optimization problem to minimize the time-average expected service delay by jointly optimizing resource allocation and task offloading while satisfying the energy constraints. To solve the formulated problem, we first develop a Lyapunov-based online control algorithm to decompose it into multiple one-slot problems. Since each one-slot problem is a large-scale mixed-integer nonlinear program (MINLP) that is intractable for classical computers, we further propose novel hybrid quantum-classical generalized Benders' decomposition (HQCGBD) algorithms to solve the problem efficiently by leveraging quantum advantages in parallel computing. Numerical results validate the effectiveness of the proposed MEC-enabled SATIN schemes. 
\end{abstract}



\section{Introduction} \label{sec:intro}  

With the proliferation of Internet of Things (IoT) devices, global mobile data traffic is estimated to surge by a factor of 3.5, reaching 329 \si{EB} per month in 2028 \cite{mobiledata}. The huge influx of data will cause an enormous burden on traditional cloud computing networks. Besides, the emerging artificial intelligence applications of IoT devices, such as face recognition and real-time video analytics, are typically latency-critical and computation-intensive. Traditional cloud computing systems face difficulties in meeting these demands. By pushing network control, computing, and storage to the network edges (e.g., access points and cellular base stations), multi-access edge computing (MEC) is considered a promising paradigm to reduce network congestion and provide low-latency computation service \cite{ding2018beef}. However, developing terrestrial infrastructures to provide MEC service in remote areas is still infeasible. 

Recently, integrating the terrestrial network with the non-terrestrial network has been regarded as a critical technique to increase network coverage and capacity \cite{giordani2020non, zhang2024energy, yu2020joint}. Low Earth orbit (LEO) satellites play important roles in non-terrestrial networks. With orbital heights ranging from 500\si{km} to 2,000 \si{km}, LEO satellites can offer several advantages, including lower development costs and lower service delays compared with the geosynchronous Earth orbit (GEO) and medium Earth orbit (MEO) satellites. Another key component of non-terrestrial networks is the high-altitude platform (HAP), which offers a promising wireless solution for complementing and enhancing the existing satellite and terrestrial networks. Operating in the stratosphere (from 17 to 22 \si{km}), HAPs have fewer geographical restrictions than terrestrial base stations and supply higher network throughput than LEO satellites. Therefore, HAPs can serve as aerial base stations to improve the quality of service (QoS) and reduce deployment costs. 
%
%
By integrating both LEOs and HAPs with terrestrial base stations, the satellite-aerial-terrestrial integrated network (SATIN) offers a comprehensive and versatile solution to tackle diverse communication challenges and paves the way for the upcoming 6G cellular network \cite{giordani2020non}. {\color{black} Most existing works on SATINs focus on the communication service while ignoring the computation service \cite{alsharoa2020improvement, liu2023resource, jia2021toward}. Some recent studies \cite{ding2021joint, mei2022energy} have started to investigate MEC-enabled SATINs. However, in these studies, the researchers mainly focus on optimizing the system latency and/or energy consumption in a fixed satellite network. Their solution cannot be directly applied to MEC-enabled SATINs, since the channel condition and connectivity of satellites are time-varying. Only a few recent studies \cite{waqar2022computation, gong2022computation, zhang2020satellite} investigate dynamic resource allocation and task offloading in MEC-enabled SATINs. However, these studies either focus on the computation capacity of HAPs while neglecting the computation capacity of satellites \cite{waqar2022computation} or only consider a single satellite with computation capacity \cite{gong2022computation, zhang2020satellite}.} Moreover, in these studies, the SATIN optimization problems are formulated as mixed-integer nonlinear programs (MINLPs), which are NP-hard. In large-scale real-world SATIN applications, it is challenging to leverage classical computing techniques to obtain optimal solutions.

To overcome this challenge, quantum computing (QC) has emerged as a new promising approach for solving combinatorial optimization problems \cite{bharti2022noisy}. Unlike classical computing, which processes information using binary bits, QC utilizes qubits to encode the superposition of states so that QC can explore exponential combinations of states simultaneously. This feature enables QC to solve large-scale real-world optimization problems more efficiently and faster. There are two major paradigms in QC: gate-based QC and adiabatic QC (AQC). Gate-based QC uses discrete quantum gate operations to manipulate qubits, achieving the desired final state after evolution. The primary limitation of gate-based QC is that the depth of the circuit and the number of qubits are limited. There is no efficient gated-based QC optimization algorithm available for industrial applications. For instance, we can only obtain less than 150 qubits for gate-based QC from IBM \cite{IBM}. Unlike gate-based QC, AQC encodes the problem into the Hamiltonian of the quantum system, whose ground states induce optimal solutions. AQC is hard to implement due to the susceptibility of quantum physical systems to non-ideal conditions. Quantum annealing (QA) can be regarded as a relaxed AQC that does not necessarily require universality or adiabaticity \cite{QA_adv}. QA is typically implemented in single-instruction machines called quantum annealers. Currently, more than 5,000 qubits are available for QA from D-wave \cite{dwave}. With a large number of qubits, QA has the potential to solve real-world problems such as car manufacturing scheduling \cite{QA_car}, RNA folding \cite{QA_RNA1,QA_RNA2}, satellite beam placement \cite{dinh2023efficient}, and robust fitting optimization \cite{doan2022hybrid}.


{\color{black}In this paper, we propose the first QA approach for improving service provisioning in dynamic MEC-enabled SATINs. Particularly, we take into account the unique features of LEO satellites and HAPs (i.e., mobility and/or limited energy) and formulate a stochastic program under uncertainty. Then, we propose a joint communication/computation resource allocation and task offloading scheme to achieve service delay minimization. By exploiting the Lyapunov optimization approach, we design an online algorithm to decouple the $T$-slot problem into multiple one-slot problems. Since each one-slot problem is a large-scale MINLP that is generally NP-hard and intractable for classical computing, we further propose several hybrid quantum-classical generalized Benders’ decomposition (HQCGBD) algorithms to solve it efficiently.} Our main contributions are summarized as follows.
\begin{itemize}
    \item We formulate a novel optimization problem for jointly optimizing resource allocation and task offloading with the goal of minimizing time-average expected service delay in MEC-enabled SATINs. 
    \item We propose a Lyapunov-based online optimization approach to transfer the original problem into multiple one-slot problems, which can be solved without requiring future information.
    %
    %
    \item As each one-slot problem is a large-scale MINLP, which is generally intractable to solve by classical computers, we develop a hybrid quantum-classical algorithm named HQCGBD by integrating generalized Benders' decomposition (GBD) and QA to solve the problem efficiently. Furthermore, inspired by the parallel processing capability of QC, we further design a multi-cut strategy to accelerate the convergence of HQCGBD.
    \item We conduct extensive experiments to evaluate the proposed framework and algorithms. The results highlight that our proposed hybrid quantum-classical computing algorithms outperform the classical computing algorithm. Furthermore, the proposed MEC-enabled SATIN scheme demonstrates its superiority over various baselines.
\end{itemize}

The remainder of this paper is organized as follows. The related works are discussed in Section \ref{Sec:related_works}. The preliminaries are introduced in Section \ref{sec:bg}. The system model and problem formulation are described in Section \ref{sec:model}. In Section \ref{sec:online}, we present the online control algorithm to solve the formulated problem. Simulation results are presented in Section \ref{sec:numerical}. Finally, Section \ref{sec:conclusion} concludes the paper.

\section{Related Works}\label{Sec:related_works}
In this section, we discuss the most related prior works from two aspects: SATIN system and quantum annealing.

\subsection{Resource Management for SATINs}

SATIN systems have attracted considerable research interest in recent years due to their potential benefits in advancing 5G and 6G networks \cite{giordani2020non}. There is various literature on resource management in SATIN that aims at optimizing network throughput \cite{alsharoa2020improvement, liu2023resource, jia2021toward, gong2022computation, zhang2020satellite}, energy consumption\cite{ding2021joint, mei2022energy}, and overall cost of system latency and/or energy consumption \cite{waqar2022computation}. Most prior works in the area of SATIN (e.g., \cite{alsharoa2020improvement, liu2023resource, jia2021toward}) ignore the computing capability provided by satellites and HAPs and mainly focus on their communication aspect. Alsharoa et al. \cite{alsharoa2020improvement} studied the joint resource allocations and the HAPs’ locations problem aiming at maximizing the users’ throughput. Liu et al. \cite{liu2023resource} maximized the total throughput of the secondary network for the NOMA-enabled cognitive SATIN by jointly optimizing transmission power and subchannel allocation. Based on the time expanding graph (TEG), Jia et al. \cite{jia2021toward} jointly optimized resource allocation and data flow aiming at maximizing the total throughput. Only a few studies \cite{ding2021joint,mei2022energy} start to consider computing with satellites’ and HAPs' on-board resources. Ding et al. \cite{ding2021joint} optimized user association, multi-user multiple input and multiple output (MU-MIMO) transmit precoding, computation task assignment, and resource allocation to minimize the energy consumption of SATIN. Mei et al. \cite{mei2022energy} formulated a computation task offloading and resource allocation optimization problem to minimize the system energy consumption. However, in these studies, the focus is primarily on optimizing latency and/or energy in static satellite networks, which is not suitable for dynamic SATIN in practice. Recent studies \cite{gong2022computation, zhang2020satellite, waqar2022computation} start to consider MEC in dynamic SATIN. Waqar et al. \cite{waqar2022computation} formulated a joint computation offloading and resource allocation optimization problem in the dynamic SATIN with MEC to minimize the task latency and energy consumption cost.
Gong et al. \cite{gong2022computation} proposed a dynamic three-layer SATIN model to maximize the network throughput by jointly optimizing the communication and computation resources. Zhang et al. \cite{zhang2020satellite} studied the problem of joint two-tier user association and offloading decisions aiming at maximizing the network throughput. However, most works in the SATIN systems formulate their problems as MINLPs, which are hard for classical computers to solve. Therefore, they either utilize heuristics or complex optimization techniques to tackle them. Inspired by QC, we propose an HQCGBD algorithm to solve these MINLPs efficiently.

\subsection{Quantum Annealing for Optimization}
\label{subsec:QA_for_Opti}
Extensive research efforts have been made recently to utilize QA for solving practical optimization problems \cite{fan2022hybrid}. However, the major limitation of QA is that it only accepts quadratic unconstrained binary optimization (QUBO) formulation. To solve the continuous optimization problem, we need to utilize a large number of ancillary qubits to discretize continuous variables, which is costly. Owing to this reason, most prior studies typically formulate real-world applications as binary quadratic model (BQM) problems  \cite{QA_car,QA_RNA1,QA_RNA2} or mixed-integer linear programming (MILP) problems \cite{ doan2022hybrid,dinh2023efficient}. These formulations are advantageous as they can be readily transformed into the QUBO format, making them suitable for efficient optimization with QA. Only a limited number of recent studies \cite{zhang2024quantum, zhang2024quantum_1} have commenced leveraging QA to address formulated MINLP problems. {\color{black}For example, Zhang et al. \cite{zhang2024quantum} leverage QA to solve the formulated MINLP with the goal of maximizing network throughput by jointly optimizing the content delivery policy, cache placement, and transmission power allocation in an integrated satellite-terrestrial network.  However, these formulations do not adequately capture the dynamic nature of SATIN systems. To the best of our knowledge, this is the first work to integrate QA and Lynapnova optimization in solving a stochastic optimization problem aimed at optimizing service delay for MEC-enabled SATINs.}


\section{Preliminaries} \label{sec:bg}
In this section, we first introduce the background of QA, followed by a concise overview of the QA workflow.
\subsection{Theoretical Background of QA}

QA  has been designed to solve classical large-scale combinatorial optimization problems \cite{QA_car,QA_RNA1,QA_RNA2, doan2022hybrid,dinh2023efficient}. These optimization problems often involve minimizing a cost function, which can be equated to finding the ground state of a classical Ising Hamiltonian $H_p$ \cite{lucas2014ising}. However, many formulated optimization problems have numerous local minima, corresponding to Ising Hamiltonians that are reminiscent of classical spin glasses. Owing to the abundant local minima, it is challenging for traditional algorithms to obtain the global minimum \cite{kadowaki2002study}. QA emerges as a potent alternative for tackling these complex tasks. It transforms the classical Ising Hamiltonian $H_p$ into the quantum domain, representing it as a collection of interacting qubits.

Based on the adiabatic theorem of quantum mechanics \cite{bharti2022noisy}, we can initialize the quantum system’s state in the ground state of some initial Hamiltonian $H_0$, which is known and easy to prepare. Then, we gradually evolve the system’s state to the targeted Hamiltonian $H_P$, ensuring that the system consistently stays in the ground state throughout the evaluation period. By measuring the final ground state of the system, we can obtain the ground state of targeted Hamiltonian $H_p$, which is also the solution to the original optimization problem \cite{yarkoni2022quantum}. We denote the evolution time as $\tau_e \in [0, T_e]$ and the number of qubits as $K$. This time-dependent evolution can be  written as 
\begin{equation} \label{eq: quantum}
    H(\tau_e) = A(\tau_e)H_0 + B(\tau_e)H_p,
\end{equation}
where $H_0=\sum_{i = 1}^K\sigma^x_i, H_p=\sum_{i  = 1}^Kh_i\sigma^z_i + \sum_{i,j =1}^KJ_{i,j}\sigma^z_i\sigma^z_j.$
Here, $A(\tau_e)$ and $B(\tau_e)$ are the annealing path functions, which are monotonic with $A(0)=1, A(T_e)=0$ and $B(0)=0, B(T_e)=1$. $h_i$ and $J_{i,j}$ are system parameters called bias and coupling strength, respectively. $\sigma_i^x$ is the $x$-Pauli matrix, while $\sigma_i^z$ is the $z$-Pauli matrix, acting on the $i$-th qubit \cite{lucas2014ising}. From Equation \eqref{eq: quantum}, we can see that the contribution of the initial Hamiltonian $H_0$ is slowly reduced while the magnitude of the targeted Hamiltonian $H_P$ is increased. At the end of the evolution, we can obtain $H(T_e)=H_p$, from which the global optimal solution can be derived.

Based on \cite{das2005quantum}, we can replace the quantum $\sigma^z$ Pauli operators with classical spin variables in Hamiltonian $H_p$ and obtain an Ising model representing Hamiltonian $H_p$, i.e,
\begin{equation} \label{eq: Ising}
\min_{\textbf{s} \in \{-1,1\}^K} H_p(\textbf{s})=\sum_{i=1}^Kh_is_i + \sum_{i=1}^{K}\sum_{j=1}^{K}J_{i,j}s_is_j.
\end{equation} 
Alternatively, we can express the optimization problem as a QUBO formulation. Let $f_Q:\{0,1\}^K \rightarrow \mathbb{R}$ be a quadratic polynomial over binary variables $\mathbf{x}=[x_1,\dots,x_K]$, and $\mathbf{Q}\in \mathbb{R}^{K\times K}$ be an upper triangular matrix. The QUBO formulation is given as
\begin{equation} \label{QUBO_1}
\min_{\textbf{x}\in \{0,1\}^K}f_Q(\mathbf{x})=\sum_{i=1}^KQ_{ii}x_i + \sum_{i=1}^{K}\sum_{j=1}^{K}Q_{ij}x_ix_j=\mathbf{x}^\intercal\mathbf{Q}\mathbf{x}.
\end{equation}
Note that the QUBO formulation can be easily transformed back into the Ising model by mapping $x_i = \frac{s_i+1}{2}$.

Although we have the theoretical guarantees of the adiabatic theorem, maintaining adiabaticity is challenging in practice due to open quantum systems' vulnerability to background noise and thermal fluctuations. Moreover, the required annealing time is proportional to the spectral energy gap between the ground and the first excited state, which is an unknown prior \cite{kumar2018quantum}. Due to these reasons, the system's ground state may be disrupted during the evolution, potentially yielding a result that is not the global optimal solution. To address this challenge, QA is performed multiple times to enhance the probability of identifying high-quality solutions in practice.


\begin{figure}[t!]
\centering
\includegraphics[scale=0.40]{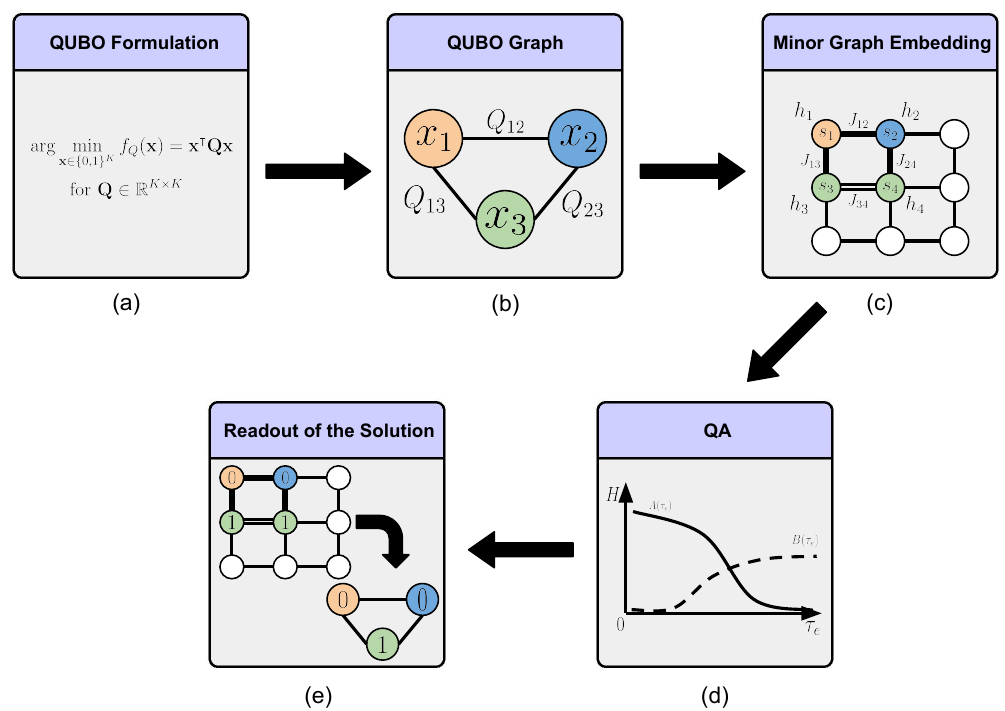}
\caption{Overview of QA workflow on a quantum annealer platform.}
\label{fig:dwave-system}
\vspace{-10pt}
\end{figure}

\subsection{QA Workflow}

As illustrated in Fig.~\ref{fig:dwave-system}, the QA workflow consists of five steps \cite{yarkoni2022quantum} as follows: \textbf{(a) QUBO formulation}: We need to formulate the real-world application problem into the QUBO formulation (i.e., Equation \eqref{QUBO_1}), which is the standard input format for quantum annealers; \textbf{(b) QUBO graph}: The quantum system then converts the QUBO formulation into a QUBO graph; \textbf{(c) Minor graph embedding}: Since the QUBO graph may not be directly compatible with the physical topology of the quantum processing unit (QPU), the quantum system finds a minor embedding of the problem graph that aligns with the sparse native topology of the QPU. In the minor embedding, each logical node may be mapped into multiple physical qubits. Those qubits are coupled with sufficient strong interactions; \textbf{(d) QA}: According to predefined annealing functions, the quantum system evolves from the initial to the targeted Hamiltonian to minimize energy; \textbf{(e) Readout of the solution}: At the end of the QA process, the qubits are either in an eigenstate or a superposition of eigenstates in the computational basis. Each eigenstate represents a potential minimum of the final Hamiltonian. The quantum system reads the individual spin values of the qubits as the candidate solution to the original problem.

\section{System Model and Problem Formulation} \label{sec:model}
In this section, we first introduce the SATIN system model. After that, the detailed task model and task execution model are described. Finally, we formulate an optimization problem to minimize the time-average expected service delay. Table. \ref{table: notation} is the summary of the notations.

\subsection{Network Model} \label{subsec:network model}

\begin{figure}[t!]
\centering
\includegraphics[scale=0.18]{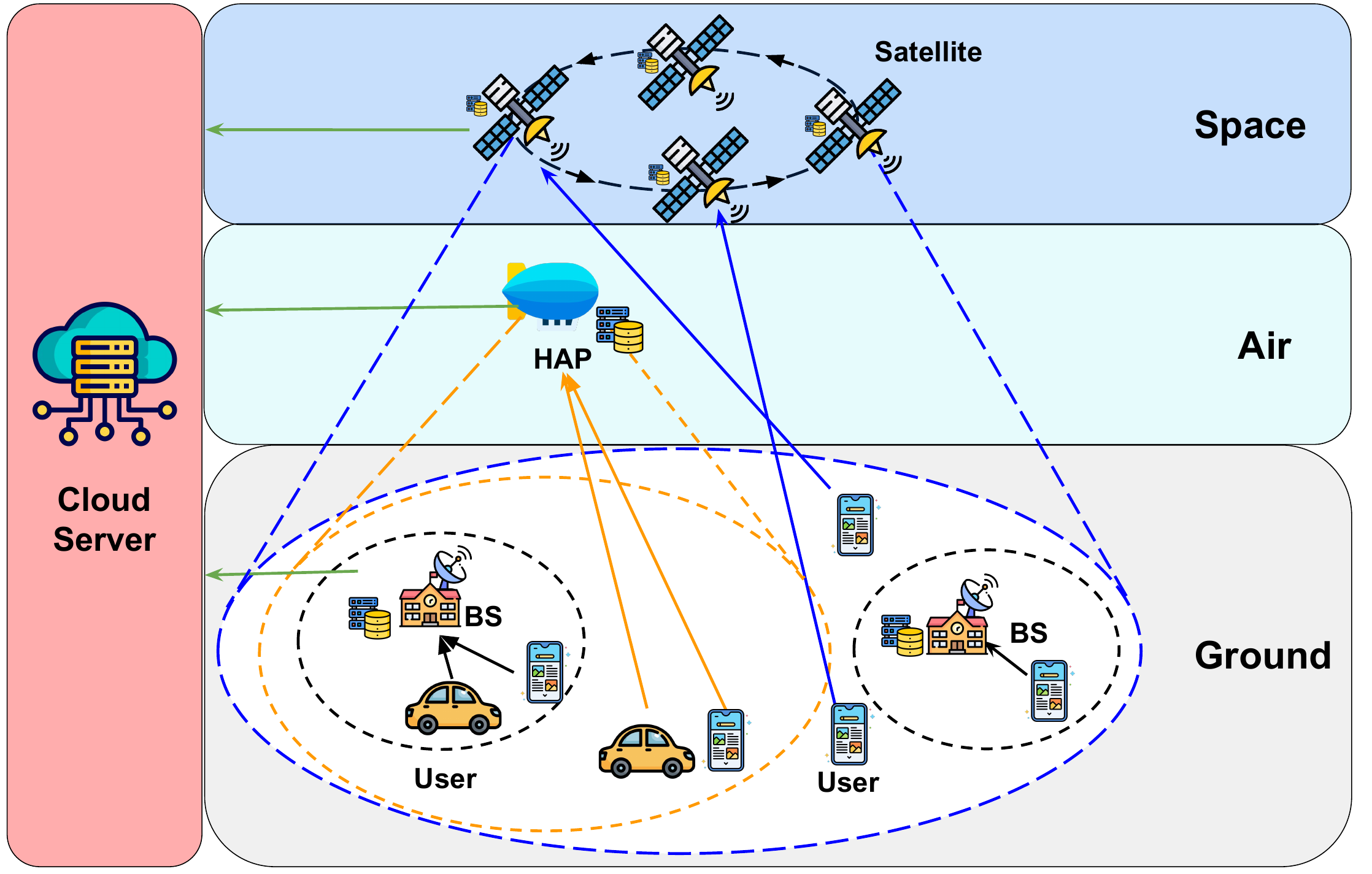}
\caption{System architecture of a MEC-enabled SATIN system.}
\label{fig:system}
\end{figure}

As shown in Fig.~\ref{fig:system}, we consider a MEC-enabled SATIN consisting of a set of users $u \in \mathcal{U} := \{1,\dots, U\}$ and a set of access points (APs) $m \in \mathcal{M} := \{1,\dots, M\}$. The APs are divided into three tiers: 
\begin{enumerate*}
  \item ground tier with a set of base stations (BSs) $m \in \mathcal{B} := \{1,\dots, B\}$;
  \item air tier with a set of high-altitude platforms (HAPs) $m \in \mathcal{H} := \{B+1,\dots, H+B\}$; and
  \item space tier with a set of satellites $m \in \mathcal{S} := \{H+B+1,\dots, M\}$. 
\end{enumerate*}
Satellites, BSs, and HAPs are deployed to provide users with both relaying and computation services through the satellite-to-user (S2U), BS-to-user (B2U), and HAP-to-user (H2U) channels \footnote{Throughout this work, we assume the handover process between APs can be managed by the central control server and leave the design of a more sophisticated handover process to the future work.}, respectively. Similar to \cite{han2021joint, hassan2022seamless,tang2021computation}, we assume that all APs are connected to a cloud server $c$ via backhaul links \footnote{With the rapid development of communication technology, it is possible for different APs to connect with the same cloud server using various backbone networks \cite{fidler2010optical, toyoshima2008ground, qian2020edge}.}. Without loss of generality, we assume the cloud server $c$ is equipped with sufficient computation capabilities. There is a central network controller equipped with classical and quantum computers to manage the task offloading and resource allocation for the SATIN. On the other hand, each AP is equipped with an edge server, which has limited computation capabilities. For ease of exposition, we divide the time horizon $T$ into discrete time slots with slot duration $\tau$, which is indexed by $t \in \mathcal{T}:=\{1,\dots, T\}$. 


\begin{table}[t!] 
\caption{List of Notations}
\label{table: notation}
\centering
\begin{adjustbox}{width=0.85\columnwidth,center}
\begin{tabular}{|C{1.2cm} | C{5.6cm} |}
\hline
\textbf{Notation} & \textbf{Definitions}  \\
\hline
$\mathcal{U}$/$\mathcal{M}$/$\mathcal{B}$/ $\mathcal{H}$/$\mathcal{S}$/$\mathcal{T}$ & Users set/APs set/BSs set/HAPs set/satellites set/time slots set\\
\hline
$D_{u}^t$ & Task input data size of user $u$  \\
\hline
$C_{u}^t$ & Number of CPU cycles to process 1-bit of task input data of user $u$ \\
\hline
$A_{u,m}^t$ & Available connection between user $u$ and AP $m$\\
\hline
$\alpha_{u,m}^t$ & Association between user $u$ and AP $m$\\
\hline
$z_{u, m}^t$ & Computation offloading decision between user $u$ and AP $m$\\
\hline
$r_{u,m}^t$ & Transmission data rate from user $u$ to AP $m$\\ 
\hline
$\beta_{u,m}^t$ & Fraction of bandwidth allocated to user $u$ by AP $m$\\
\hline
$\tau_{u,m}^{\text{TX},t}$ & Transmission delay for offloading computing task from user $u$ to AP $m$ \\
\hline
$\tau_{u,m}^{\text{CP},t}$ & Computation delay for processing the task of user $m$ by AP $m$\\ 
\hline
$\tau_{u,m,c}^{\text{TX},t}$ & Transmission delay for relaying the computation task of user $u$ from AP $m$ to the cloud server\\
\hline
$\tau_{u,m,c}^{\text{CP},t}$ & Transmission delay of AP $m$ for transmitting task of user $u$ to the cloud server \\
\hline
$\Bar{e}_m$ & Energy budget of AP $m$\\
\hline
$e^{\text{TX},t}_{u,m,c}$ & Energy consumption of AP $m$ for transmitting task of user $u$ to the cloud server \\
\hline
$e_{u,m}^{\text{CP},t}$ & Computation energy consumption for processing the task of user $u$ at AP $m$ \\
\hline
$f_{u,m}^t$ & Computation resources allocated to process the task of user $u$ by AP $m$\\
\hline
\end{tabular}
\end{adjustbox}
\vspace{-10pt}
\end{table}

\subsection{Task Model}

Similar to \cite{guo2020online, jiang2022joint}, the tasks considered in this paper are sequential dependent, which means a new task is generated after the previous task is processed. For instance, in firefighting or underwater exploration scenarios, a mobile robot periodically conducts the simultaneous localization and mapping (SLAM) task to sense the environment, generate a navigation map, identify its current position, and track its movements \cite{yang2019multi}. These SLAM tasks must be completed in real time, especially for life-or-death applications. Based on this, we assume each user $u$ generates a computation-intensive task at the beginning of time slot $t$ and completes this task in the same time slot by offloading it to the associated AP. 
%
The computation task of user $u$ is modeled as a tuple $\mathcal{W}_{u}^t := (D_{u}^t, C_{u}^t)$, where $D_{u}^t$ (in bits) denotes the task input data size, and $C_{u}^t$ (in CPU cycles/bit) denotes the number of CPU cycles to process 1-bit of task input data.

\subsection{Task Execution Model} \label{subsec:task_e_m}

Considering the randomness of the wireless environment (e.g., the existence of an obstacle), we model the connectivity between user $u$ and AP $m$ by a discrete-time random process $A_{u,m}^t$, which indicates whether there is an available connection between user $u$ and AP $m$ at time slot $t$, i.e., 
\begin{equation}
  A_{u, m}^t =
    \begin{cases}
      1 &  \text{if user}~u~\text{can connect to AP}~m~\text{at time $t$},\\
      0 & \text{otherwise}.
    \end{cases}       
\end{equation} 

At the beginning of each time slot $t$, users first check the connectivity of all APs. Then, each user $u$ offloads its computation task $\mathcal{W}_{u}^t$ to one of the available APs. Let $\alpha_{u,m}^t \in \{0, 1\}$ denote the association between user $u$ and AP $m$, where $\alpha_{u,m}^t  = 1$ indicates user $u$ is associated with AP $m$ at time slot $t$, and otherwise $\alpha_{u,m}^t  = 0$. There are several constraints the user association decisions must satisfy. First, user $u$ can associate with AP $m$ only if AP $m$ is available at time slot $t$, which is represented as
\begin{equation} \label{const: connectivity}
\alpha_{u,m}^t \leq A_{u, m}^t, \quad \forall u, m, t.
\end{equation}
Second, considering each user $u$ can be associated with only one AP simultaneously at any time slot $t$, we have the following constraint:
\begin{equation} \label{const: association}
\sum_{m \in \mathcal{M}}\alpha_{u,m}^t = 1,\quad \forall u, t.
\end{equation}

After receiving the entire input data of task $\mathcal{W}_{u}^t$, the associated AP $m$ will determine whether the task should be processed locally or further offloaded to the cloud server. Let $ z_{u, m}^t \in \{0, 1\}$ denote the computation offloading decision. $z_{u, m}^t = 1$ indicates the task $\mathcal{W}_{u}^t$ is processed in AP $m$ at time slot $t$. Otherwise, the task is offloaded to the cloud for processing, and $z_{u, m}^t=0$. Since AP $m$ processes the task of user $u$ only when the user is associated with the AP at time slot $t$, we have the following constraint:
\begin{equation} \label{const: connectivity_vs_computing}
\alpha_{u,m}^t \geq z_{u, m}^t,\quad \forall u, m, t.
\end{equation}

In the following, we model the energy consumption and delay incurred during the communication and computation procedures. 

\subsubsection{Communication Model}

While the Doppler effects need to be considered in the communication model for SATINs, thanks to recent developing techniques \cite{nieto2018doppler, ali2005doppler}, the Doppler shifts are assumed to be estimated and compensated accurately. Besides, similar to \cite{alsharoa2020improvement}, we assume that the orthogonal frequency division multiple access (OFDMA) is adopted, and there is neither intra-cell nor inter-cell interference. Based on these assumptions, we model the channel gain from user $u$ to AP $m$ as a discrete-time random process $g_{u,m}^t$. According to Shannon's Theorem, the instantaneous transmission data rate from user $u$ to AP $m$ at time $t$ is expressed as
\begin{equation}\label{eq:rate}
    r_{u,m}^t = \beta_{u,m}^tB_{m}\log_2\left(1 + \frac{P_{u}g_{u,m}^t}{\sigma^2}\right), 
\end{equation}
where $P_u$ is the transmission power of user $u$, $\sigma^2$ is the noise variance, $B_{m}$ is the total bandwidth of AP $m$, and $\beta_{u,m}^t \in [0, 1]$ is the fraction of bandwidth allocated to user $u$ by AP $m$. Since an AP only allocates bandwidth to its connected users, we have the following constraint:
\begin{equation}  \label{const: speed_vs_association}
    \beta^{\min}_m \alpha_{u,m}^t \leq \beta_{u,m}^t \leq \beta^{\max}_m \alpha_{u,m}^t,\quad \forall u, m, t,
\end{equation}
where $\beta^{\min}_m$ and $\beta^{\max}_m$ are the minimum and maximum bandwidth fraction of AP $m$ allocated for each associated user, respectively.
%
%
Moreover, the sum of allocated bandwidth cannot exceed the total bandwidth of AP $m$. Therefore, we have the following constraint on $\beta_{u,m}^t$, i.e.,
\begin{equation} \label{const: allocated_bandwidth}
    \sum_{u \in \mathcal{U}}\beta_{u,m}^t \leq 1,\quad \forall m, t.
\end{equation}
Given the data rate \eqref{eq:rate}, the transmission delay $\tau_{u,m}^{\text{TX},t}$ for offloading computing task $\mathcal{W}_{u}^t$ from user $u$ to AP $m$ at time $t$ should satisfy the following:
\begin{equation} \label{const: tx_u_m}
    \tau_{u,m}^{\text{TX},t} r_{u,m}^t \geq \alpha_{u,m}^t D_{u}^t,\quad \forall u, m, t. 
\end{equation}

After an AP receives the computation task, the AP will either process it or relay it to the cloud server. The transmission delay for relaying the computation task of user $u$ from AP $m$ to the cloud server $c$ at time slot $t$ satisfies the following:
\begin{equation} \label{const: tx_u_m_c}
    \tau_{u,m,c}^{\text{TX},t}r_{m,c} \geq \alpha_{u,m}^t(1-z_{u,m}^t)D_{u}^t,\quad \forall u, m, t,
\end{equation}
where $r_{m,c}$ is the pre-set backhaul data rate between AP $m$ and cloud server $c$. Accordingly, the energy consumption of AP $m$ for transmitting task $\mathcal{W}_{u}^t$ to the cloud server $c$ at time slot $t$ is calculated as
\begin{equation} \label{const: power_u_m_c}
e^{\text{TX},t}_{u,m,c}=P_m\tau_{u,m,c}^{\text{TX},t},\quad \forall u, m, t,
\end{equation}
where $P_m$ is the transmission power of AP $m$.

\subsubsection{Computation Model} At each time slot $t$, the APs will decide whether the offloaded tasks are processed locally or further offloaded to the cloud server for processing. If the task of user $u$ is processed locally, the associated AP $m$ will allocate computation resources (i.e, CPU frequency) $f_{u,m}^t$ to process the task, which satisfies the following:
\begin{equation} \label{const: computing_vs_resource}
    f^{\min}_m z_{u,m}^t \leq f_{u,m}^t \leq f_m^{\max} z_{u,m}^t,\quad \forall u, m, t
\end{equation}
where $f_m^{\min}$ and $f_m^{\max}$ are the minimum and maximum allocated computation resources of AP $m$ for each task, respectively. Moreover, each AP has a limit on its maximum CPU frequency modeled as:  
\begin{equation} \label{const: allocated_computing_resource}
\sum_{u \in \mathcal{U}}f_{u,m}^t \leq F_m, \quad \forall  m, t,
\end{equation}
where $F_m$ is the maximum computation capability of AP $m$. %
The corresponding computation delay $\tau_{u,m}^{\text{CP}, t}$ for the task $\mathcal{W}_{u}^t$ at AP $m$ in time $t$ should satisfy the following:
\begin{equation} \label{const: computing_u_m}
    \tau_{u,m}^{\text{CP},t} f_{u,m}^t \geq  z_{u, m}^tD_{u}^tC_{u}^t,\quad \forall u, m, t.
\end{equation}
According to \cite{wang2016mobile}, we model the computation power of AP $m$ at time slot $t$ as $\kappa_m(f_{u,m}^t)^3$, where $\kappa_m$ is the effective switched capacitance depending on the CPU architecture of AP $m$. Then,  the corresponding computation energy consumption for processing the task $\mathcal{W}_{u}^t$ at AP $m$ in time $t$ is given by
\begin{equation} \label{const: e_computing_u_m}
    e_{u,m}^{\text{CP},t} = \kappa_m(f_{u,m}^t)^3\tau_{u,m}^{\text{CP},t},\quad \forall u, m, t.
\end{equation}

On the other hand, if AP $m$ does not process the task $\mathcal{W}_{u}^t$ onboard, AP $m$ will offload the task to the cloud server. The computation delay $\tau_{u,m,c}^{\text{CP},t}$ of the task $\mathcal{W}_{u}^t$ at cloud server $c$ is given by
\begin{align}  \label{const: computing_u_m_c}
    &\tau_{u,m,c}^{\text{CP},t} F_{c} \geq  (1-z_{u,m}^t)\alpha_{u,m}^t D_{u}^t C_{u}^t,\quad \forall u, m, t,
\end{align}
where $F_{c}$ represents the pre-assigned CPU frequency of the cloud server for each user. 

Based on the above models, the total service delay of task $\mathcal{W}_{u}^t$ at time slot $t$ is given by
\begin{equation} 
    O_{u}^t = \sum_{m \in \mathcal{M}}(\tau_{u,m}^{\text{TX},t} + \tau_{u,m}^{\text{CP},t} + \tau_{u,m,c}^{\text{TX},t}+ \tau_{u,m,c}^{\text{CP},t}).
\end{equation}
Besides, the total energy consumption of AP $m$ to process the offloaded task of user $u$ at time slot $t$ is given by
\begin{equation} 
    e_{u,m}^{\text{AP},t} =  e^{\text{TX},t}_{u,m,c}+e_{u,m}^{\text{CP},t}.
\end{equation}


Since APs, particularly HAPs and LEO satellites, are usually power-limited and the energy consumption of AP $m$ depends on the random connectivity $A_{u,m}^t$ and channel gain $g_{u,m}^t$, we consider the following energy constraint on the time-average expected energy consumption of each AP $m$:
\begin{align}
    \frac{1}{T}\sum_{t \in \mathcal{T}}\sum_{u \in \mathcal{U}} \mathbb{E}\{ e_{u,m}^{\text{AP},t}\} &\leq \Bar{e}_m \label{const:expect_m}, \quad \forall m, 
\end{align} 
where $\Bar{e}_m$ is the energy budget of AP $m$. 
 
\subsection{Problem Formulation}

In this paper, we are interested in minimizing the time-average expected service delay over a large time horizon. Therefore, the control problem can be stated as follows: for the dynamic SATIN system, design a control strategy which, given the past and the present random connectivity and channel gain, chooses the task offloading decisions $\bm{\alpha}^t = \{\alpha_{u, m}^t\}$, computation decision $\mathbf{z}^t  = \{z_{u, m}^t\}$, communication resource allocation $\bm{\beta}^t  =\{\beta_{u, m}^t\}$, computation resource allocation $\mathbf{f}^t =\{f_{u, m}^t\}$, and processing delay $\bm{\tau}^t =\{\tau_{u,m}^{\text{TX},t},\tau_{u,m}^{\text{CP},t}, \tau_{u,m,c}^{\text{TX},t}, \tau_{u,m,c}^{\text{CP},t}\}$ such that the time-average expected service delay is minimized. It can be formulated as the following stochastic optimization problem:
\begin{subequations}
\begin{flalign} \label{P0}
\textbf{P}_0: \min_{\{\bm{\alpha}^t, \mathbf{z}^t , \bm{\beta}^t, \mathbf{f}^t, \bm{\tau}^t\}} \quad & \frac{1}{T}\sum_{t \in \mathcal{T}} \sum_{u \in \mathcal{U}} \mathbb{E}\{O_u^t\} \\
\text{s.t.} \quad 
\quad & \bm{\alpha}^t, \mathbf{z}^t \in \{0, 1\}^{U\times M}, \quad \forall t \label{constr:binary}\\
\quad & \bm{\beta}^t, \mathbf{f}^t,  \bm{\tau}^t\geq \bm{0}, \quad \forall t \label{const:non-negative}\\
\quad &\eqref{const: connectivity} \text{-}\eqref{const: connectivity_vs_computing},\eqref{const: speed_vs_association}\text{-}\eqref{const: tx_u_m_c}, \notag \\
\quad & \eqref{const: computing_vs_resource}\text{-}\eqref{const: computing_u_m},\eqref{const: computing_u_m_c}, \eqref{const:expect_m}.
\end{flalign}
\end{subequations}

 One challenge of solving this optimization problem lies in the uncertainty of connectivity and channel state information, which makes problem $\textbf{P}_0$ stochastic. Another challenge is the energy constraint \eqref{const:expect_m} brings the ``time-coupling property" to problem $\textbf{P}_0$. In other words, the current control action may impact future control actions, making the problem $\textbf{P}_0$ more challenging to solve. Moreover, the continuous decision variables $\{\bm{\beta}^t, \mathbf{f}^t, \bm{\tau}^t\}$ are tightly coupled with binary decision variables $\{\bm{\alpha}^t, \mathbf{z}^t\}$, which makes problem $\textbf{P}_0$ a large-scale MINLP that is NP-hard.

\section{Online Control Algorithm Design} \label{sec:online}

In this section, we design an online control algorithm for the central network controller to solve problem $\textbf{P}_0$. We first utilize a Lyapunov-based optimization approach to decompose the $T$-slot problem into multiple one-slot problems. Since each one-slot problem is still a large-scale MINLP, which is intractable, we then leverage GBD to decompose each one-slot problem into a master problem and a subproblem. Although we can transform the non-convex subproblem into a convex problem and efficiently solve it with classical computers, the master problem is a mixed-integer linear program (MILP). Inspired by QC, we convert the master problem into the QUBO formulation and use QA to solve the reformulated master problem efficiently. Furthermore, based on the powerful parallel computing capabilities of quantum computers, we further propose a specialized quantum multi-cut strategy to speed up the optimization process. 

\subsection{Lyapunov-Based Optimization Approach} \label{subsec:lyapunov}
Equivalently, we can transform the constraint \eqref{const:expect_m} into a queue stability constraint \cite{guo2012optimal, guo2012electricity,guo2013decentralized, guo2013energy}. In detail, we first construct a virtual energy consumption queue $Q_m^t$, which represents the backlog of energy consumption for AP $m$ at the current time slot $t$. The updating equation of queue $Q_m^t$ is given by
\begin{equation}
    Q_m^{t+1}:=  \max{\left\{Q_m^t+\sum_{u \in \mathcal{U}}e_{u,m}^{\text{AP},t}- \Bar{e}_m, 0\right\}}, \quad \forall m, t. \label{ap_queue}
\end{equation}
We can easily show that the stability of virtual queue \eqref{ap_queue} ensures the constraint \eqref{const:expect_m}. Then, we define a quadratic Lyapunov function as
$
    L(t):= \frac{1}{2} \sum_{m \in \mathcal{M}}(Q_m^t)^2.
$
For ease of presentation, we define $\bm{Q}(t)=\{ Q_m^t\}_{\forall m}$ at time slot $t$. Therefore, the one-slot conditional Lyapunov drift can be described as $\Delta(t) := \mathbb{E}\left\{L(t+1)-L(t)|\bm{Q}(t)\right\}.$
%
%
Next, we define the following Lyapunov \emph{drift-plus-penalty} term:
\begin{align}
    \Delta_V(t) &:= \Delta(t) + V\mathbb{E}\left\{\sum_{u \in \mathcal{U}}O_u^t|\bm{Q}(t)\right\},
\end{align}
where $V$ is a non-negative parameter that controls the trade-off between the optimality of objective function and stability of queue backlogs. We have the following lemma regarding the \emph{drift-plus-penalty} term. 

\begin{lemma}
\emph{Under any feasible action that can be implemented at time slot $t$, we have}
\begin{align} \label{lyapunov_upper}
    \Delta_V(t) \leq& C^*+ V\mathbb{E}\left\{\sum_{u \in \mathcal{U}}O_u^t|\bm{Q}(t)\right\} \notag\\
        &+\sum_{m \in \mathcal{M}}\mathbb{E}\left\{Q_m^t\left(\sum_{u \in \mathcal{U}}e_{u,m}^{\text{AP},t}- \Bar{e}_m\right)|\bm{Q}(t)\right\},
\end{align}
where $C^*= (1/2)\sum_{m \in \mathcal{M}}((\sum_{u \in \mathcal{U}}E_{u,m}^{\text{AP},t})^2 + \Bar{e}_m^2)$ is a constant value over all time slots. Here $E_{u,m}^{\text{AP},t} =  \max{\left\{P_m D_u / r_{m,c}, \kappa_m(f_m^{\max})^2D_{u}^tC_{u}^t\right\}}$. 

\end{lemma}
\begin{IEEEproof}
Please see the Appendix A.
\end{IEEEproof}

Now, we present our online control algorithm. The main idea of our algorithm is to choose control actions that minimize the R.H.S. of \eqref{lyapunov_upper}. We first initialize $\bm{Q}(0)=\bm{0}$. At each time slot $t$, the cloud server collects the current channel gain $g_{u,m}^t$ and connectivity $A_{u, m}^t, \forall u, m$, and do:
\begin{enumerate}[left=0pt]
    \item Choose the control decisions $\bm{\alpha}^t$, $\mathbf{z}^t$, $\bm{\beta}^t$, $\mathbf{f}^t$, and $\bm{\tau}^t$ as the optimal solution to the following optimization problem:
    \begin{flalign*} \label{P1}
    \textbf{P}_1: \min_{\bm{\alpha}^t, \mathbf{z}^t, \bm{\beta}^t, \mathbf{f}^t,  \bm{\tau}^t} \quad & \Phi(\bm{\alpha}^t, \mathbf{z}^t, \mathbf{f}^t,  \bm{\tau}^t)\\
    \text{s.t.} 
    \quad &\eqref{const: connectivity}\text{-}\eqref{const: connectivity_vs_computing},\eqref{const: speed_vs_association}\text{-}\eqref{const: tx_u_m_c},\\
\quad & \eqref{const: computing_vs_resource}\text{-}\eqref{const: computing_u_m},\eqref{const: computing_u_m_c},\eqref{constr:binary},\eqref{const:non-negative},
    \end{flalign*}
    where $\Phi(\bm{\alpha}^t, \mathbf{z}^t, \mathbf{f}^t,  \bm{\tau}^t) = V\sum_{u \in \mathcal{U}}O_u^t
     +\sum_{m \in \mathcal{M}}Q_m^t($ $\sum_{u \in \mathcal{U}}e_{u,m}^{\text{AP},t} - \Bar{e}_m).
    $
    \item Update $\bm{Q}(t)$ according to the dynamics \eqref{ap_queue}. 
\end{enumerate}

Then, We analyze the performance of the proposed Lyapunov-based online optimization approach when the connectivity $A^{t}_{u,m}, \forall t$ and channel gain $g^{t}_{u,m}, \forall t$ are IID stochastic processes. Note that our results can be extended to the more general setting where $A^{t}_{u,m}$ and $g^{t}_{u,m}, \forall t$ evolves according to some finite state irreducible and aperiodic Markov chains, according to the Lyapunov optimization framework \cite{guo2012optimal, guo2012electricity,guo2013decentralized, guo2013energy}:
\begin{theorem}
\emph{If $A^{t}_{u,m}$ and $g^{t}_{u,m}, \forall t$ are IID over time slot $t$, then the time-average expected objective value under our algorithm is within bound $C^*/V$ of the optimal value, i.e.,}
\begin{align}
\mathop{\textup{lim sup}}_{T\rightarrow\infty} \frac{1}{T}\sum_{t \in \mathcal{T}} \sum_{u \in \mathcal{U}} \mathbb{E}\{O_u^t\} \leq \textbf{P}_0^* + \frac{C^*}{V},
\end{align}
\emph{where $\textbf{P}_0^*$ is the optimal objective value, $C^*$ is the constant given in Lemma 1, and $V$ is a control parameter}.
\end{theorem}

\begin{IEEEproof}
Please see the Appendix B.
\end{IEEEproof}

Note that problem $\textbf{P}_1$ is MINLP, which is generally intractable for classical computers. Moreover, if we utilize a large number of ancillary qubits to discretize continuous variables and solve it directly by QA, the cost is extremely high. In the following subsection, we design a hybrid quantum-classical approach to solving problem $\textbf{P}_1$. For simplicity of notation, we will omit the superscript $t$ without causing ambiguity in the following.

\subsection{Hybrid Quantum-classical Generalized Benders’ Decomposition} \label{subsec:HQCGBD}

As the binary decision variables $\{\bm{\alpha}$, $\mathbf{z}\}$ are coupled with the continuous decision variables $\{\bm{\beta}, \mathbf{f},  \bm{\tau}\}$, the optimal solution of problem $\textbf{P}_1$ is not easily obtained. We adopt the GBD to solve this MINLP. Particularly, we first decompose problem $\textbf{P}_1$ into two problems, a subproblem and a master problem. On the one hand, the subproblem is a non-convex optimization problem with continuous decision variables $\{\bm{\beta}, \mathbf{f},  \bm{\tau}\}$ when binary decision variables $\{\bm{\alpha}$, $\mathbf{z}\}$ are fixed. It can be transformed into an equivalent convex form and solved by classical computers. The solution of the subproblem provides an upper bound for the optimal value of problem $\textbf{P}_1$. On the other hand, the master problem is a MILP when continuous decision variables $\{\bm{\beta}, \mathbf{f},  \bm{\tau}\}$ are fixed. We propose a novel quantum optimization approach to solve it with respect to binary decision variables $\{\bm{\alpha}$, $\mathbf{z}\}$, and obtain a lower bound for the optimal value of problem $\textbf{P}_1$. The subproblem and master problems are iteratively solved until the upper and lower bounds converge. We describe the detailed solution procedures of the subproblem and master problem in the following.      
 
\subsubsection{Classical Optimization for Subproblem}
Given the fixed binary decision variables $\bm{\alpha}^{(l)}$ and $\mathbf{z}^{(l)}$ obtained from the master problem at the $(l-1)$-th iteration, the subproblem can be written as
\begin{flalign*}
\textbf{SP}_1: \min_{\bm{\beta}, \mathbf{f},  \bm{\tau}} \quad &  \Phi(\bm{\alpha}^{(l)}, \mathbf{z}^{(l)}, \mathbf{f},  \bm{\tau}) \\
\text{s.t.} 
\quad & \eqref{const: speed_vs_association}\text{-}\eqref{const: tx_u_m_c}, \eqref{const: computing_vs_resource} \text{-} \eqref{const: computing_u_m},\eqref{const: computing_u_m_c}, \eqref{const:non-negative}.
\end{flalign*}
We can observe that constraints \eqref{const: tx_u_m} and \eqref{const: computing_u_m} are still non-convex due to the product terms $\tau_{u,m}^{\text{TX}} r_{u,m}$ and $\tau_{u,m}^{\text{CP}} f_{u,m}$, respectively. For non-convex term $\tau_{u,m}^{\text{TX}} r_{u,m}$,  note that AP $m$ will allocate at least $\beta_m^{\min}$ bandwidth fraction to each connected user. We can divide both sides of \eqref{const: tx_u_m} by non-zero $r_{u,m}$ for the connected users. The resulting equation is convex. Similarly, we can divide both sides of \eqref{const: computing_u_m} by non-zero $f_{u,m}$ for the onboard computing tasks. For simplicity of notation, we denote the associated users with AP $m$ as $\mathcal{U}_{1,m}^{(l)} \in \mathcal{U}$ (i.e., $\alpha_{u,m}^{(l)}=1, \forall u \in \mathcal{U}_{1,m}^{(l)}$) and the users whose task is processed in AP $m$ as $\mathcal{U}_{2,m}^{(l)} \in \mathcal{U}$ (i.e., $z_{u,m}^{(l)}=1, \forall u \in \mathcal{U}_{2,m}^{(l)}$). Then, the reformulated subproblem $\textbf{SP}_1$ can be written as 
\begin{subequations}
\begin{flalign} 
\textbf{SP}_2: \min_{\bm{\beta}, \mathbf{f},  \bm{\tau}} ~ &  \Phi(\bm{\alpha}^{(l)}, \mathbf{z}^{(l)}, \mathbf{f},  \bm{\tau}) \\
\text{s.t.} 
~ & \tau_{u,m}^{\text{TX}} \geq \frac{\alpha_{u,m}^{(l)}D_{u}}{r_{u,m}},~ \forall u\in \mathcal{U}_{1,m}^{(l)}, m \in \mathcal{M} \label{sp2_1}\\
 ~ & \tau_{u,m}^{\text{CP}} \geq  \frac{z_{u, m}^{(l)}D_{u}C_{u}}{f_{u,m}},~ \forall u\in \mathcal{U}_{2,m}^{(l)}, m \in \mathcal{M}\\
~ & \eqref{const: speed_vs_association},\eqref{const: allocated_bandwidth}, \eqref{const: tx_u_m_c}, \eqref{const: computing_vs_resource},\eqref{const: allocated_computing_resource},\eqref{const: computing_u_m_c}, \eqref{const:non-negative}. \label{sp2_3}
\end{flalign}
\end{subequations}

Subproblem $\textbf{SP}_2$ is convex as it only has linear objective function and convex constraints. Solving its dual problem is equivalent to solving subproblem $\textbf{SP}_2$. We formulate its dual problem as  
\begin{flalign} \label{dual}
\max_{\bm{\xi}}~\min_{\bm{\beta}, \mathbf{f},  \bm{\tau}} ~\mathcal{L}(\bm{\alpha}^{(l)}, \mathbf{z}^{(l)}, \bm{\beta}, \mathbf{f},  \bm{\tau}, \bm{\xi})
\end{flalign}
Here, $\mathcal{L}$ is the Largangian function of subproblem $\textbf{SP}_2$, $\{\bm{\beta}, \mathbf{f},  \bm{\tau}\}$ are the primary variables, and $\bm{\xi}$ are the dual variables associated with constraints $\eqref{sp2_1}-\eqref{sp2_3}$. Please see Appendix C for the details of the dual problem. 

Due to the convexity of subproblem $\textbf{SP}_2$, it can be efficiently solved by the classical numerical solvers (e.g., Mosek \cite{aps2022mosek}). After we obtain both the primary and dual solutions of subproblem $\textbf{SP}_2$, we can utilize them to generate a \emph{Benders' cut} as input to the master problem. Specifically, the optimal objective value of subproblem $\textbf{SP}_2$ is the upper bound of the optimal objective value of problem $\textbf{P}_1$ because subproblem $\textbf{SP}_2$ is a constrained version of problem $\textbf{P}_1$ where all binary decision variables are fixed.

\subsubsection{Quantum Optimization for Master Problem} \label{subsec: QO_for_MP}
Based on the optimal solutions of subproblem $\textbf{SP}_2$ at the $l$-th iteration, the master problem is defined as 
\begin{subequations}\label{mp}
\begin{flalign} 
\textbf{MP}_1: \min_{\bm{\alpha}, \textbf{z}, \mu} \quad & \mu  \label{master_obj}\\
\text{s.t.} 
\quad & \mu \geq \mathcal{L}(\bm{\alpha}, \mathbf{z}, \bm{\beta}^{(k)}, \mathbf{f}^{(k)},  \bm{\tau}^{(k)}, \bm{\xi}^{(k)}) , \notag\\
\quad & \qquad\forall k \in \{1, \dots,l\}\label{benders_cut}\\
\quad & \eqref{const: connectivity}\text{-}\eqref{const: connectivity_vs_computing},\eqref{constr:binary},
\end{flalign}
\end{subequations}
where constraints \eqref{benders_cut} are the \emph{Benders' cuts}, and $\mu$ is a slack variable. In each iteration, we introduce a new cut, generated from subproblem $\textbf{SP}_2$, into the master problem. Therefore, the search space for the globally optimal solution is gradually narrowed down. Moreover, the optimal value of $\mu$ can be regarded as the performance lower bound of problem $\textbf{P}_1$. %
Note that master problem $\textbf{MP}_1$ is still a large-scale MILP, which is hard to solve for classical computers. Besides, different from the QUBO formulation \eqref{QUBO_1}, master problem $\textbf{MP}_1$ contains a continuous variable as the objective function and many constraints. Thus, we utilize the following steps to equivalentl transform it into the QUBO formulation so that it can be solved by QA \cite{zhao2022hybrid, zhao2023optimal}.

\emph{Objective function reformulation:} Note that master problem $\textbf{MP}_1$ contains continuous variable $\mu$, while QA only accepts binary variables as input. Thus, we utilize a binary vector $\textbf{w}$ with a length of $N$ bits to approximate continuous variable $\mu$ and denote it as $\Bar{\mu}(\textbf{w})$:
\begin{align} \label{QUBO:master}
    \Bar{\mu}(\mathbf{w}) = & \sum_{i=0}^{n_1}w_{i}2^{i-\underline{n}_+}w_{i}  -\sum_{j=n_2}^{N}w_{j}2^{j-n_2}w_{j},
\end{align}
where $n_1 = \overline{n}_++\underline{n}_+$, $n_2 = 1+ \overline{n}_++\underline{n}_+$, and $N = 1+\overline{n}_++\underline{n}_++\overline{n}_-$.Here, $\overline{n}_+, \underline{n}_+,\overline{n}_-$ are the number of bits representing the positive integer, positive decimal, and negative integer part of $\mu$, respectively. Remarkably, in the objective function tailored for the QUBO formulation, we only need to discretize a single continuous variable. Note that increasing the number of binary variables for discretization enhances the precision of representation but also increases representation costs. In practice, we first estimate the range of values for the continuous variable. Then, based on the precision requirements, we determine the number of binary variables. This approach significantly reduces computational cost compared to directly discretizing all continuous variables in the original problem.

\begin{algorithm}[t]
\caption{Proposed HQCGBD Algorithm} 
\label{alg:1}
\begin{algorithmic}[1]
\REQUIRE Initialize $\bm{\alpha}^{(0)}$, $\mathbf{z}^{(0)}$, $\text{UB}^{(0)} = +\infty$, and $\text{LB}^{(0)}=-\infty$. Set the iteration index $l=1$, the maximum iteration number $L^{\text{max}}$, and a small constant $\epsilon \rightarrow 0$. 
\WHILE{$l < L^{\text{max}}$ or $|\frac{\text{UB}^{(l-1)}-\text{LB}^{(l-1)}}{\text{UB}^{(l-1)}}| > \epsilon$}
        \STATE Solve subproblem $\textbf{SP}_2$ with the fixed binary decision variables $\bm{\alpha}^{(l-1)}$ and $\mathbf{z}^{(l-1)}$ in the classical computer. 
        \STATE Calculate the Benders' cut according to \eqref{dual} and then add it to the master problem $\textbf{MP}_1$.
        \STATE Update $\text{UB}^{(l)} = \min\{\text{UB}^{(l-1)},\Phi^{(l)}\}$.
    %
        \STATE Transform master problem $\textbf{MP}_1$ into its QUBO formulation $\textbf{MP}_2$ according to \eqref{QUBO:master}-\eqref{QUBO4}.
        \STATE Solve the problem $\textbf{MP}_2$ by D-Wave’s quantum  annealer.
        \STATE Obtain the optimal solution $\bm{\alpha}^{(l)}$, $\mathbf{z}^{(l)}$, and $\mu^{(l)}$. 
        \STATE  Update $\text{LB}^{(l)}=\mu^{(l)}$ and $l$ = $l$ + 1.
\ENDWHILE 
\ENSURE Optimal $\bm{\alpha}^*, \mathbf{z}^*, \bm{\beta}^*, \mathbf{f}^*,  \bm{\tau}^*$.
\end{algorithmic}
\end{algorithm}
\vspace{-10pt}

\emph{Constraints reformulation:} After we reformulate the objective function of master problem $\textbf{MP}_1$, an ILP master problem can be obtained. However, the reformulated master problem is still constrained, which makes it not directly applicable for QA. According to the constraint-penalty pair principle in \cite{zhao2022hybrid}, we further convert constraints \eqref{const: connectivity}-\eqref{const: connectivity_vs_computing} and \eqref{benders_cut} as: 
\begin{equation}\label{QUBO1}
f_Q^{\eqref{const: connectivity}}(\bm{\alpha}, \mathbf{s}) = \sum_{u \in \mathcal{U}}\sum_{m \in \mathcal{M}}\zeta_{1,u,m} (\alpha_{u,m} - A_{u, m} + \sum_{x=0}^{\overline{x}_{1,u,m}}2^x{s}_{1,x,u,m})^2,
\end{equation}
where $\overline{x}_{1,u,m} = \big\lceil \log_2\big(\min_{\bm{\alpha}}(
     A_{u, m}-\alpha_{u,m} )\big) \big\rceil$.
\begin{equation}\label{QUBO2}
    f_Q^{\eqref{const: association}}(\bm{\alpha}) = \sum_{u \in \mathcal{U}}\zeta_{2,u}(\sum_{m \in \mathcal{M}}\alpha_{u,m} - 1)^2.
\end{equation}
\begin{equation}\label{QUBO3}
    f_Q^{\eqref{const: connectivity_vs_computing}}(\bm{\alpha}, \textbf{z}) = \sum_{u \in \mathcal{U}}\sum_{m \in \mathcal{M}}\zeta_{3,u,m}(z_{u, m}-z_{u, m}\alpha_{u,m}).
\end{equation}
\begin{flalign}\label{QUBO4}
    f_Q^{\eqref{benders_cut}}(\bm{\alpha}, \mathbf{z},\mathbf{s}) =& \sum_{k=1}^l\zeta_{5,k}( \mathcal{L}(\bm{\alpha}, \mathbf{z}, \bm{\beta}^{(k)}, \mathbf{f}^{(k)},  \bm{\tau}^{(k)}, \bm{\xi}^{(k)}) \notag\\
    &-\Bar{\mu}(\textbf{w})+ \sum_{x=0}^{\overline{x}_{2,k}}2^x{s}_{2,x, k})^2,
\end{flalign}
where
$
\overline{x}_{2,k} = \lceil \log_2(\min_{\bm{\alpha},\mathbf{z}}(
     - \mathcal{L}(\bm{\alpha}, \mathbf{z}, \bm{\beta}^{(k)}, \mathbf{f}^{(k)},  \bm{\tau}^{(k)}, \bm{\xi}^{(k)})+ \Bar{\mu}(\textbf{w}))) \rceil.
$

Here, $\bm{s}$ is the binary slack variables, $\Bar{\bm{x}}$ is the upper bound of the number for $\bm{s}$, and $\bm{\zeta}$ is the penalty parameters which are defined according to \cite{kochenberger2014unconstrained}. Finally, we express master problem $\textbf{MP}_1$ in the QUBO formulation as
\begin{align} \label{qubo_min}
   \textbf{MP}_2: \min_{\bm{\alpha}, \mathbf{z},\mathbf{w},\mathbf{s}} \quad
    & \Bar{\mu}(\mathbf{w}) +  f_Q^{\eqref{const: connectivity}}(\bm{\alpha}, \mathbf{s}) +  f_Q^{\eqref{const: association}}(\bm{\alpha}) \nonumber\\ 
    &+ f_Q^{\eqref{const: connectivity_vs_computing}}(\bm{\alpha}, \textbf{z})+ f_Q^{\eqref{benders_cut}}(\bm{\alpha}, \mathbf{z},\mathbf{s}). 
\end{align}

\subsubsection{Overall Algorithm} \label{subsec:overall algorithm}
Based on the analysis above, the proposed HQCGBD algorithm is summarized in Algorithm \ref{alg:1}. The algorithm contains an iterative procedure. We first initialize the binary decision variables $\bm{\alpha}^{(0)}$ and $\mathbf{z}^{(0)}$ as well as other parameters. In the $l$-th iteration, we solve subproblem $\textbf{SP}_2$ in the classical computer with the fixed binary decision variables $\bm{\alpha}^{(l-1)}$ and $\textbf{z}^{(l-1)}$, which are generated by master problem $\textbf{MP}_2$ in the last iteration (Line 2). After that, we add the calculated Benders' cut to master problem $\textbf{MP}_1$ and update the upper bound $\text{UB}^{(l)}$ by the optimal objective value $\Phi^{(l)}$ of subproblem $\textbf{SP}_2$ (Lines 3--5). Next, master problem $\textbf{MP}_1$ is reformulated as its QUBO formulation $\textbf{MP}_2$ by using the appropriate penalties (Lines 6--7). Finally, we leverage D-Wave’s quantum annealer to solve master problem $\textbf{MP}_2$ and update the performance lower bound $\text{LB}^{(l)}$ (Lines 8--10). This iterative procedure stops until the approximation gap $|(\text{UB}^{(l)}-\text{LB}^{(l)})/{\text{UB}^{(l)}}|$ is within a preset threshold $\epsilon$ or the maximal iteration index $L^{\text{max}}$ is reached. Since our proposed method follows the classical GBD framework, the complexity of the proposed algorithm aligns with the classical GBD analysis \cite{geoffrion1972generalized}. However, as we will demonstrate in the following section, our experiments verify that HQCGBD outperforms the traditional GBD running in a classical computer.

\begin{algorithm}[t]
\caption{Proposed Multi-cut HQCGBD Algorithm} 
\label{alg:2}
\begin{algorithmic}[1]
\REQUIRE Initialize $\rho$ feasible values of the binary decision variables as $\mathcal{X}^{(0)}=\{\bm{\alpha}^{(0)}_i,\mathbf{z}^{(0)}_i\}_{i=1}^{\rho}$, $\text{UB}^{(0)} = +\infty$, and $\text{LB}^{(0)}=-\infty$. Set the iteration index $l=1$, the maximum iteration number $L^{\text{max}}$, and a small constant $\epsilon \rightarrow 0$. 
\WHILE{$|\frac{\text{UB}^{(l-1)}-\text{LB}^{(l-1)}}{\text{UB}^{(l-1)}}| > \epsilon$ or $l < L^{\text{max}}$}
    \FOR{$\{\bm{\alpha},\mathbf{z}\} \in \mathcal{X}^{(l-1)}$}
        \STATE Solve subproblem $\textbf{SP}_2$ with the fixed binary decision variables $\bm{\alpha}$ and $\mathbf{z}$ in the classical computer.  
        \STATE Calculate the Benders' cut according to \eqref{dual} and then add it to master problem $\textbf{MP}_1$.
        \STATE Update $\text{UB}^{(l)} = \min\{\text{UB}^{(l-1)},\Phi^{(l)}\}$.
    \ENDFOR
    %
        \STATE Transform the master problem $\textbf{MP}_1$ into its QUBO formulation $\textbf{MP}_2$ according to \eqref{QUBO:master}-\eqref{QUBO4}.
        \STATE Solve the problem $\textbf{MP}_2$ by D-Wave’s quantum annealer.
        \STATE Obtain $\rho$ feasible solutions ${\mathcal{X}}^{(l)}=\{\bm{\alpha}^{(l)}_i,\mathbf{z}^{(l)}_i\}_{i=1}^{\rho}$ and $\{\mu^{(l)}_i\}_{i=1}^{\rho}$. 
        \STATE  Update $\text{LB}^{(l)}=\min \{\mu^{(l)}_i\}_{i=1}^{\rho}$ and $l$ = $l$ + 1.
\ENDWHILE 
\ENSURE Optimal $\bm{\alpha}^*, \mathbf{z}^*, \bm{\beta}^*, \mathbf{f}^*,  \bm{\tau}^*$.
\end{algorithmic}
\end{algorithm}
\vspace{-10pt}

\subsection{Multi-cut Strategy of HQCGBD} \label{sec: multi-cut}
\begin{figure}[t]
\centering
\subfloat[]{\includegraphics[width=0.18\textwidth, height=2.5 cm]{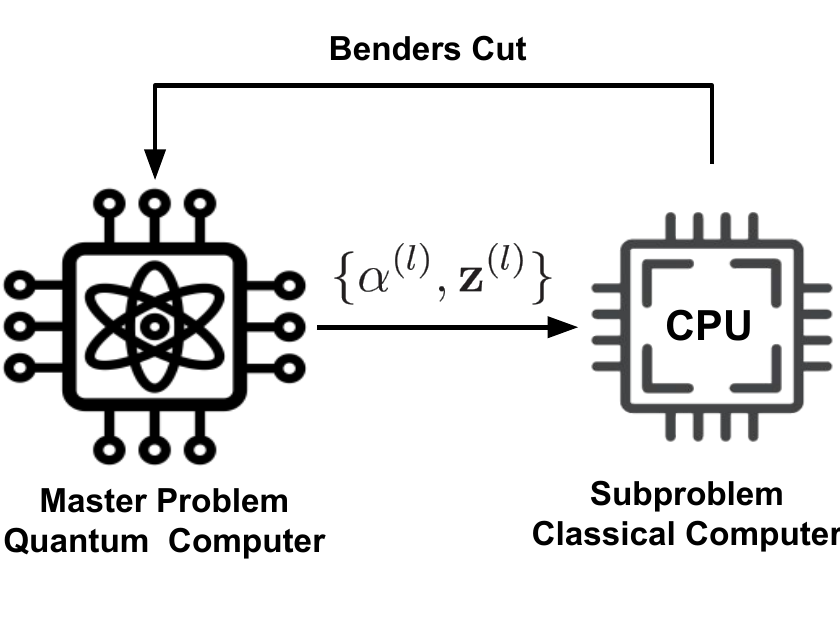}
\label{fig:cbd}}
\hspace{0.4cm}
\subfloat[]{\includegraphics[width=0.18\textwidth, height=2.5 cm]{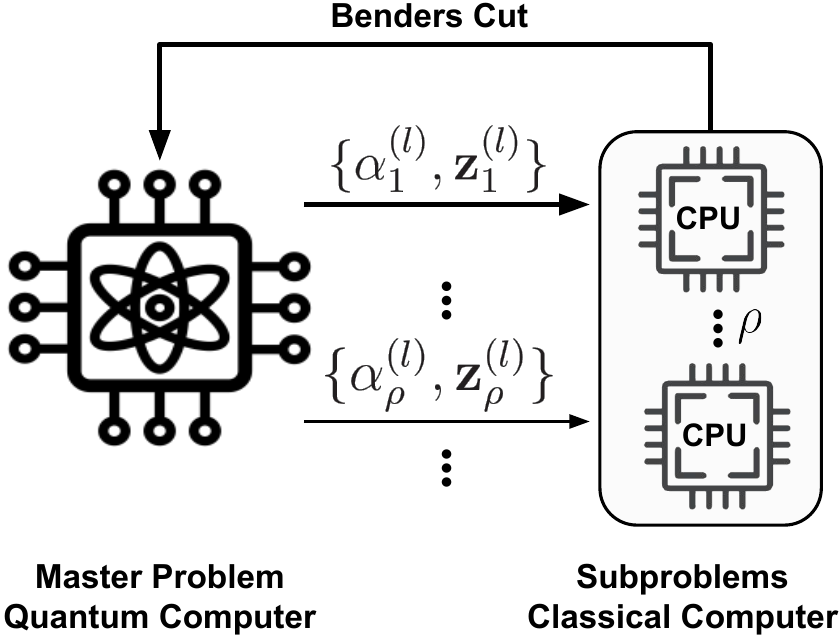}%
\label{fig:MQBC}}
\caption{An overview of (a) Single-cut HQCGBD and (b) Multi-cut HQCGBD.}
\label{fig:cbd_mqbc}
\vspace{-10pt}
\end{figure}

Even though QPU have powerful computing capacity, the implementation of single-cut HQCGBD may still require massive computing time. As shown in Fig.~\subref*{fig:cbd}, single-cut HQCGBD just generates one Benders' cut at each iteration. Single-cut HQCGBD may need numerous iterations to converge if the quality of generated Bendes' cut is low. We note that quantum computers can yield multiple feasible solutions at each iteration, which classical computers can utilize to construct multiple Benders' cuts. These cuts can improve the obtained lower bounds when solving the master problem. Based on this, we design a specialized quantum multi-cut strategy, which is shown in Fig.~\subref*{fig:MQBC}. We select top-$\rho$ feasible solutions with the lowest energies from the QA results in each iteration and utilize them as seeds to generate the multiple Benders' cuts on the classical computers for the next iteration. The detailed procedures for multi-cut HQCGBD are summarized in Algorithm \ref{alg:2}.
%

\section{Numerical Evaluation} \label{sec:numerical}
In this section, we evaluate our proposed algorithms through extensive numerical experiments. {\color{black}Due to the high cost of QPU utilization and time limitations for the developer, our experiments are limited to a small-scale setting. However, even with these hardware limitations, our results clearly demonstrate the immense potential of this technology for the future.} We implement both HQCGBD and GBD in Python 3.7. Particularly, classical MILP and convex problems are solved using Gurobi \cite{gurobi2021gurobi} and Mosek \cite{aps2022mosek}, respectively. These classical algorithms are conducted on a server equipped with a 4.2 GHz AMD Ryzen Threadripper PRO CPU and 512 GB of RAM. On the other hand, the master problem $\textbf{MP}_2$ is solved by the real-world D-Wave Advantage platform, which relies on the Pegasus topology and features more than 5000 qubits \cite{dwave}. 

\subsection{Simulation Setup} \label{subsec:setup}

We consider a service area 
%
%
where 35 ground users are uniformly distributed. At each time slot $t$, user $u$ generates a computation-intensive and latency-critical task with input data size $D_u^t \in [1,6]$ \si{Mbit}, and requires $C_u^t \in [100,500]$ \si{CPU ~cycles/bit} to process the data. We consider a MEC-enabled SATIN system to provide the computation and relaying services for the users. This MEC-enabled SATIN system consists of 4 BSs located at each vertex, 2 HAPs placed in the fixed locations at coordinates $[0.2, 0.8, 20]$\si{km} and $[0.8, 0.2, 20]$ \si{km}, respectively, and a satellite flying at altitude $780$ \si{km} with orbital velocity $4$\si{km/s}. Similar to \cite{alsharoa2020improvement}, the S2U and H2U channels are modeled as Rician channel, while the B2U channel is modeled as Rayleigh channel. Additionally, we suppose there are $500$ time slots, and the duration of each time slot is $5$\si{s}. 

For the setup of quantum computing, the continuous variable $\mu$ in master problem $\textbf{MP}_1$ is discretized by 20 binary variables. The problem was embedded into the physical QPU graph using the \emph{minor embedding} by default settings. 755 physical qubits are used. For all problems submitted to the QPU, the annealing time was set to 20 $\mu s$, and the anneal-read cycle was repeated 1000 times. The rest of our simulation parameters, unless otherwise stated, are given in Table~\ref{talbe: simulation parameters}.

\begin{table}[t!] 
\caption{Simulation Parameters}
\label{talbe: simulation parameters}
\centering
\begin{adjustbox}{width=0.85\columnwidth,center}
\begin{tabular}{|C{4.2cm}| C{3cm}|}
\hline
Parameters & Values  \\
\hline
TX power of BS/HAP/satellites/user & 41\si{dBm}/ 41\si{dBm}/42\si{dBm}/30\si{dBm} \\
\hline
Effective switched capacitance $\kappa_m$ & $10^{-28}$   \\
\hline
Antenna gain of BS/HAP/satellite & 10\si{dBi}/15\si{dBi}/50\si{dBi} \\
\hline
Carrier center frequency of BS/HAP/satellite & 5\si{GHz}/38\si{GHz}/30\si{GHz} \\
\hline
Subcarrier bandwidth of BS/HAP/satellite & 10\si{MHz}/400\si{MHz}/800\si{MHz} \\
\hline
Spectral density of noise & -174\si{dBm/Hz}\\
\hline
\end{tabular}
\end{adjustbox}
\vspace{-2pt}
\end{table}

{\color{black} Our evaluation comprises two parts. First, we assess the performance of the proposed single and multi-cut HQCGBD algorithms by comparing them with the well-known classical approach, branch-and-bound  \cite{lawler1966branch, boyd2007branch, morrison2016branch}, which solves the master problem within the GBD framework \footnote{{\color{black} Given the complexity of our problem, which involves integer programming with a vast solution space and numerous constraints, simulated annealing is not well-suited for this scenario. Nonetheless, we include an ablation study in Appendix D to compare the performance of simulated annealing with quantum annealing in smaller settings.}}. 

Next, to provide benchmarks for the performance of the proposed SATIN scheme, we compare it with the following three baselines that are proposed by some recent work \cite{zhou2019collaborative, zhou2022service,cang2023online}. 

\begin{enumerate}
    \item{\textit{Baseline 1 - Heuristic Scheme:}} In this scheme, each user is connected to the AP with the strongest signal (i.e., maximum signal-to-noise ratio), and each AP will then randomly select tasks to compute onboard and allocate the communication and computation resource to each connected user equally \cite{zhou2019collaborative}.

 
    \item{\textit{Baseline 2 - Myopic Scheme:}} This scheme neglects the energy queue backlogs and ensures that the time-average energy constraints are always satisfied in each time slot. All decisions are optimized to meet the new constraints and minimize the time-average expected service delay \cite{zhou2022service,cang2023online}.

 \item{\textit{Baseline 3 - Satellite-Only Communication Scheme:}} Similar to \cite{waqar2022computation}, satellites function solely as relays without computational capacity, while BSs and HAPs have both communication and computation capabilities.
\end{enumerate}}

\begin{figure}
\centering
  \centering
  \includegraphics[width=0.6\linewidth]{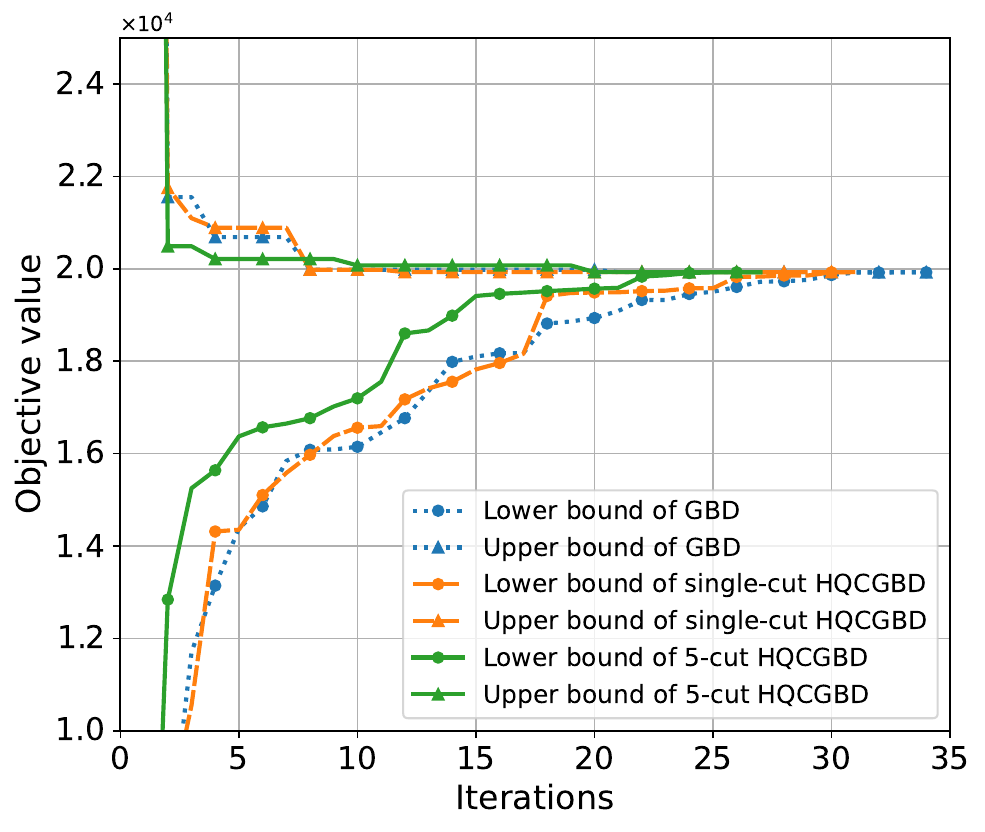}
  \caption{The objective function value for each iteration of different HQCGMBD strategies compared to the GBD approach.}
  \label{fig:convergence}
  \vspace{-10pt}
\end{figure}

\begin{figure}
\centering
\begin{minipage}[t]{0.46\columnwidth}
  \centering
  \includegraphics[width=0.93\linewidth]{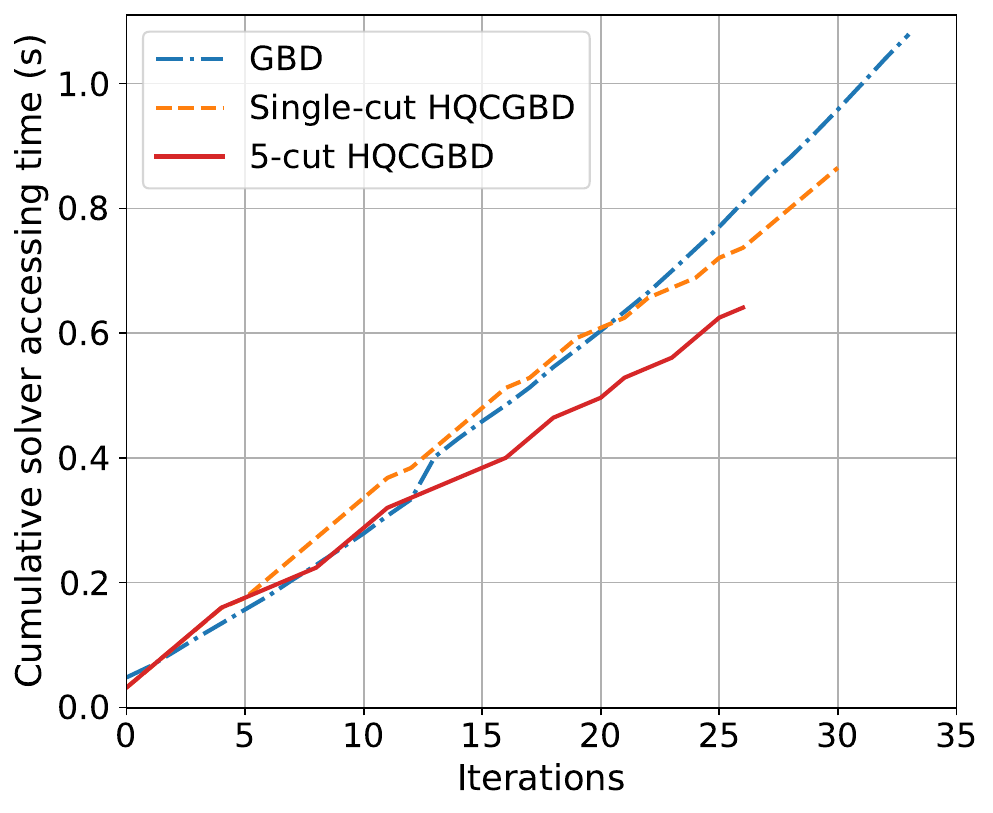}
  \caption{Cumulative solver accessing time of master problems for GBD and HQCGBDs.}
  \label{fig:time}
\end{minipage}%
\hfil
\begin{minipage}[t][2cm][t]{.22\textwidth}
  \centering
  \includegraphics[width=0.93\linewidth]{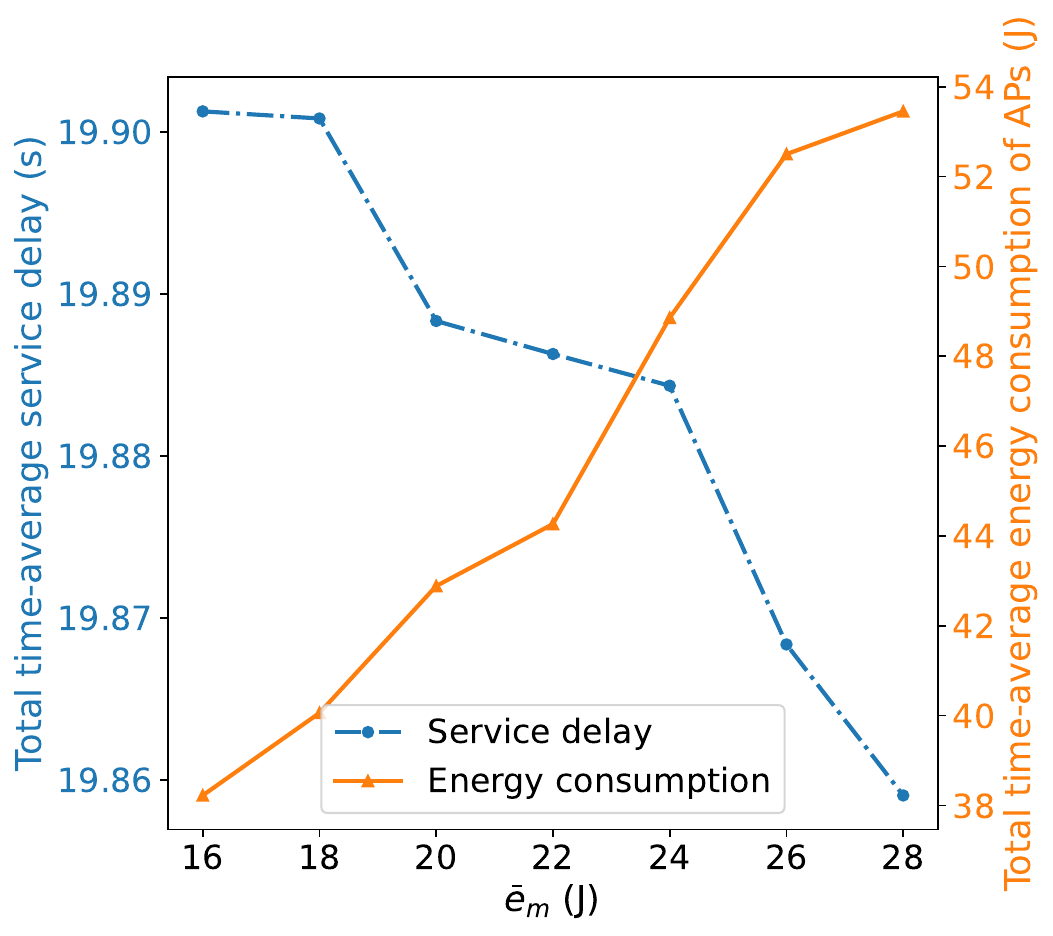}
  \captionof{figure}{Impact of $\Bar{e}_m$ on total time-average energy consumption and service delay.}
  \label{fig:impact_em}
\end{minipage}
\vspace{-10pt}
\end{figure}

\begin{figure}[t]
\centering
\subfloat[]{\includegraphics[scale=0.22]{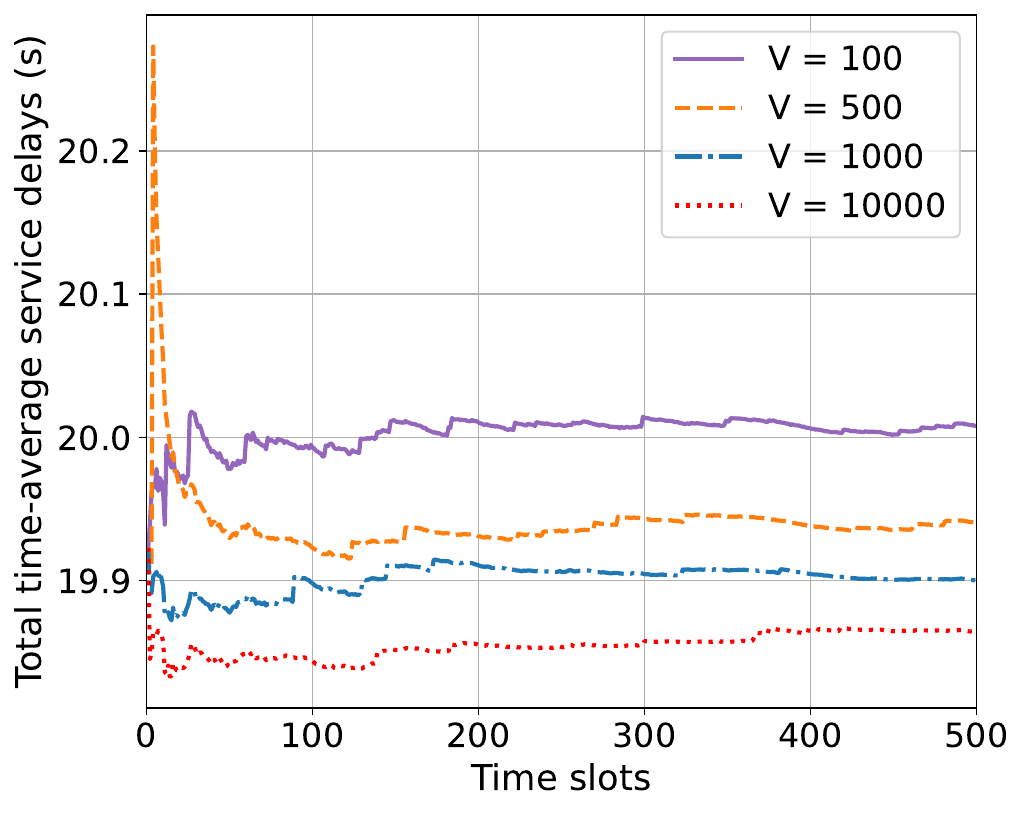}%
\label{fig:v_delay}}
\hspace{0.5cm}
\subfloat[]{\includegraphics[scale=0.22]{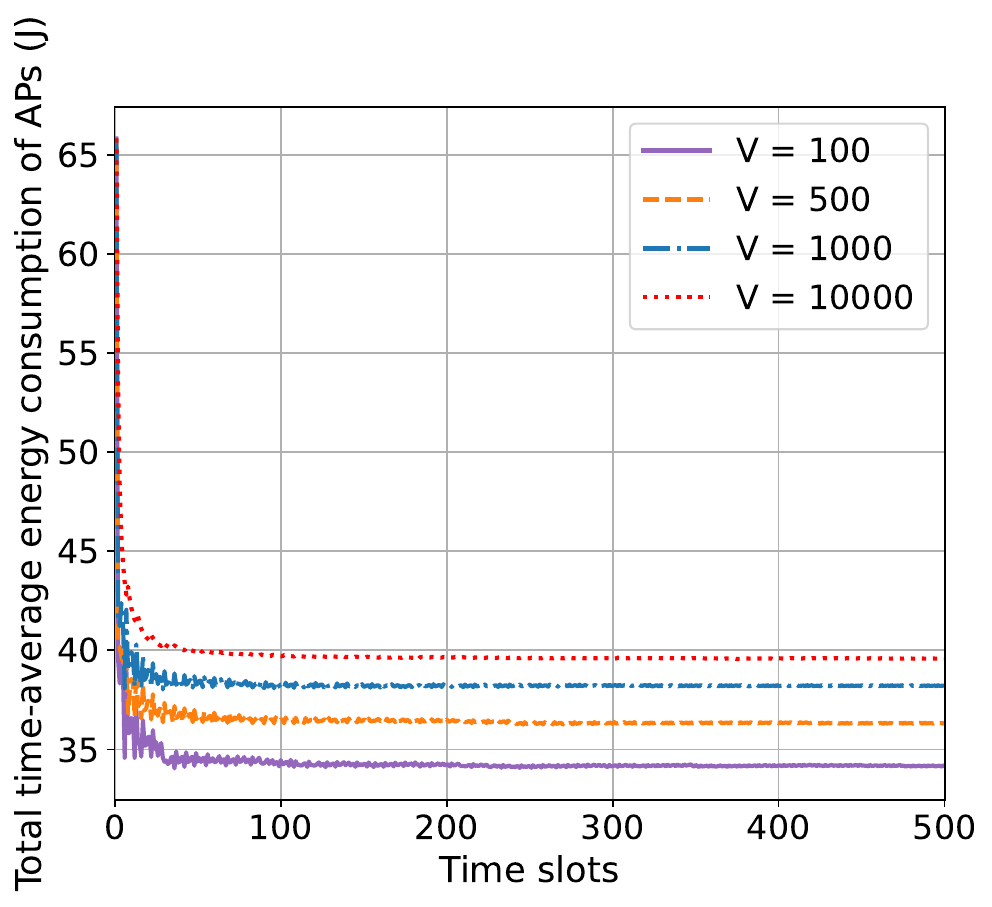}%
\label{fig:v_ap}}
\caption{Impact of V on the system performance. (a) Time-average service delays. (b) Time-average energy consumption of APs.}
\label{fig:v}
\vspace{-10pt}
\end{figure}

\begin{table}[t]
\centering
\caption{Solver accessing time of GBD and multi-cut HQCGBD strategy.}
\begin{adjustbox}{width=0.85\columnwidth,center}
\begin{tabular}{|>{\centering}p{3cm}|p{1.5cm}|p{1.5cm}|p{1cm}|}
\hline
\multirow{3}{*}{\raisebox{0.9\height}{Algorithm}}  & \multicolumn{3}{c|}{Solver Accessing Time (\si{ms})}                                                                            \\ \cline{2-4} 
                & \multicolumn{1}{>{\centering}p{1.5cm}|}{Max./Min.} & \multicolumn{1}{>{\centering}p{1.5cm}|}{Mean./Std.}  &  \multicolumn{1}{>{\centering}p{1cm}|}{Total}\\ 
                  \hline
   GBD      & \multicolumn{1}{c|}{68.29/18.10}  & \multicolumn{1}{c|}{31.74/9.34} &  \multicolumn{1}{c|}{1079.20}\\ \hline
   Single-cut HQCGBD      & \multicolumn{1}{c|}{32.11/16.02}  & \multicolumn{1}{c|}{27.91/6.99} &  \multicolumn{1}{c|}{865.08} \\ \hline
   5-cut HQCGBD      & \multicolumn{1}{c|}{32.10/16.01}  & \multicolumn{1}{c|}{23.74/7.98} &  \multicolumn{1}{c|}{640.90} \\ \hline
\end{tabular}
\end{adjustbox}
\label{tab:compare}
\end{table}

\subsection{Comparsion of Proposed HQCGBD and GBD} \label{subsec: comparsion of HQCGBD and GBD}
In this part, we first investigate the convergence properties of the proposed HQCGBD. From Fig.~\ref*{fig:convergence}, we observe that both the GBD and HQCGBD approaches can converge. 
%
%
Particularly, the GBD and single-cut HQCGBD need 34 and 31 iterations to converge, respectively. Compared with single-cut HQCGBD and GBD, the lower bound of 5-cut HQCGBD grows much faster. The reason is that the main bottleneck of GBD is the time consumed by solving the master problems, which occupies over $90 \%$ total optimization time \cite{magnanti1981accelerating}. By adopting the multi-cut strategy, we can largely improve the quality of the lower bound. Thus, 5-cut HQCGBD can reduce the number of required iterations by $23.52 \%$ compared with the GBD. 

Next, we compare the running time of GBD and our proposed algorithms. As mentioned in Section~\ref{sec: multi-cut}, the multi-cut HQCGBD involves solving multiple subproblems in each iteration. Fortunately, the complexity of each subproblem is equivalent to that of the subproblem in the GBD. Moreover, we can execute them in parallel. Hence, we only compare the performances of GBD and multi-cut HQCGBD regarding the real solver accessing time of the master problems \footnote{{\color{black}The solver accessing time is the accessing time of the quantum solver and classical solver without considering other overheads. Specifically, the access time for QPU is $T_0 \approx ((T_1 + T_2)\times T_3) + T_4$, where $T_1$ is the annealing time, $T_2$ is the readout time, $T_3$ is the number of samples, and $T_4$ is the programming time. Hence, the total QA access time $T_5$  for solving the problem (i.e., the reported QA access time) can be calculated as $T_5 = I\times T_0$, where $I$ is the HQCGBD iteration number.}}. {\color{black}From Fig.~\ref{fig:time}, we can observe that the GBD outperforms the single-cut HQCGBD before the 20-th iteration. After that, the single-cut HQCGBD performs better and better. The reason is that the master problem becomes more and more complex as we keep adding new cuts to it in each iteration. The computational time of the master problems on the quantum computers is less than that spent on the classical computers. This result demonstrates that quantum computers outperform classical computers in solving large-scale MILP problems. Specifically, the single-cut HQCGBD and 5-cut HQCGBD can save up to $19.84 \%$ and $40.61 \%$ solver accessing time of the master problem compared with the GBD, respectively. Furthermore, we show the solver accessing time of both GBD and multi-cut HQCGBD strategy in Table~\ref{tab:compare}. We can observe that the multi-cut HQCGBD strategy exhibits a consistently stable computation performance since its standard deviation of the master problem’s solver accessing time is significantly smaller than that of GBD.}

\subsection{Impact of Parameters}

\textbf{\emph{Impact of the AP's energy budget $\Bar{\bm{e}}_m$}.} To evaluate the impact of time-average energy consumption threshold $\Bar{\bm{e}}_m$, we fix the parameter $V$ and study the trade-off between the total time-average service delay and the total time-average service delay under different $\Bar{\bm{e}}_m$ in the proposed algorithm. From Fig.~\ref{fig:impact_em}, we can observe that the total time-average service delay decreases while the total time-average energy consumption of APs increases as the energy consumption threshold $\Bar{\bm{e}}_m$ of each AP $m$ increases. The reason is that AP has more energy to execute tasks onboard, eliminating the necessity of offloading tasks to the cloud server. 

\textbf{\emph{Impact of the control parameter $V$}.} Now, we focus on the trade-off between total time-average service delay and energy consumption of APs in the proposed algorithms. We choose different values of $V$ and observe the corresponding total time-average service delays and energy consumption of APs. Fig.~\ref{fig:v} shows that the total time-average service delay decreases while the total time-average energy consumption of APs increases with the increase of $V$. This is due to the fact that with the increase of $V$, our algorithm would be more aggressively minimizing the total time-average service delay, which causes larger total time-average energy consumption of APs.


\begin{figure}[t]
\centering
\subfloat[]{\includegraphics[scale=0.22]{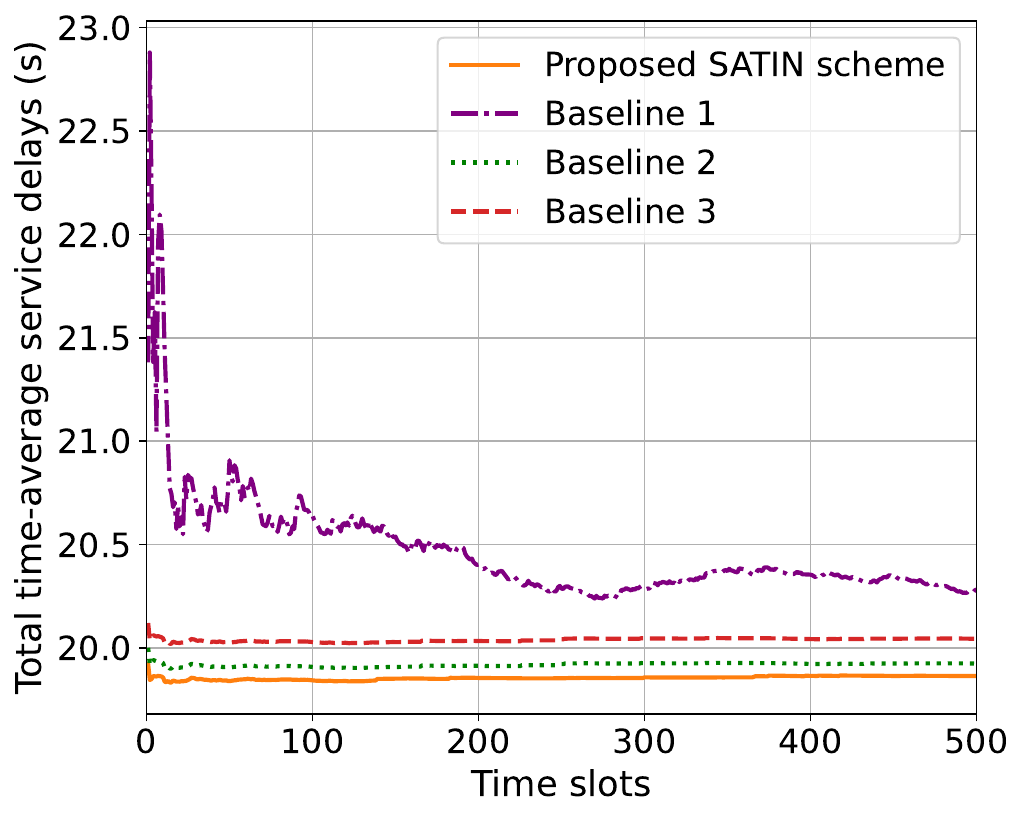}%
\label{fig:delay_bs}}
\hspace{0.5cm}
\subfloat[]{\includegraphics[scale=0.22]{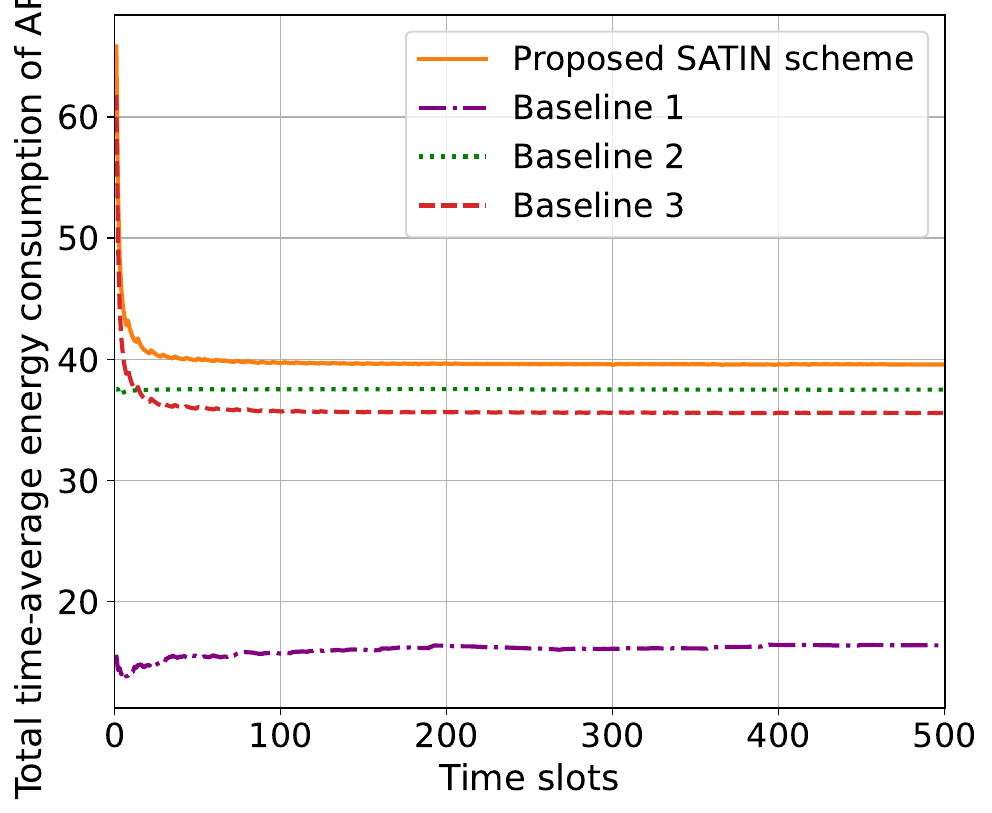}%
\label{fig:energy_bs}}
\caption{System performance comparison of different schemes. (a) Total time-average service delay. (b) Total time-average energy consumption of APs. %
}
\label{fig:bs}
\vspace{-10pt}
\end{figure}

\subsection{Advantages of Proposed Scheme} \label{sebsec:advantages}
{\color{black}In this part, we compare the performance of our proposed SATIN scheme with the baselines regarding the total time-average service delay and energy consumption of APs. From Fig.~\ref*{fig:bs}, we can observe that our proposed scheme achieves the lowest total time-average service delay while closely following the time-average energy consumption constraint. Even though the energy consumptions of other baselines are lower than the proposed SATIN scheme, their total time-average service delays are much higher, which cannot meet the service requirement of latency-critical tasks in practice.}    

\section{Conclusion} \label{sec:conclusion}
In this paper, we have investigated the joint task offloading and resource allocation problem in MEC-Enabled SATIN to minimize the time-average expected service delay. Considering the stochastic environment, a Lyapunov-based approach has been proposed to make asymptotically optimal control decisions under uncertainty, which involves solving an optimization problem at each time slot. %
%
%
%
Since each one-slot problem is a large-scale MINLP, we have developed HQCGBD to solve it. Moreover, a specialized quantum multi-cut strategy has been designed to speed up the HQCGBD convergence. Extensive simulations show the advantages of our proposed multi-cut HQCGBD in terms of iteration number until convergence and solver accessing time, while ensuring optimality. This work is our first attempt to leverage quantum computing techniques for optimizing service delay in the MEC-enabled SATIN system. Since the proposed algorithm can efficiently address large-scale MINLPs, it holds promise for various SATIN applications, e.g., routing and scheduling optimization problems. With the rapid development of quantum computers and increasing qubits \cite{dell}, we believe that quantum-assisted optimization can play an important role in the SATIN field.

\bibliographystyle{IEEEtran}
\bibliography{IEEEabrv,ref.bib}

\end{document}